\documentclass[a4paper,twocolumn,english,prl,showpacs,superscriptaddress]{revtex4}
\usepackage{mathpazo}

\usepackage[T1]{fontenc}
\usepackage[latin9]{inputenc}
\usepackage{fancyhdr}
\usepackage{babel}
\usepackage{amsmath}
\usepackage{amsthm}
\usepackage{amssymb}
\usepackage{graphicx}
\usepackage[unicode=true,pdfusetitle,
 bookmarks=true,bookmarksnumbered=false,bookmarksopen=false,
 breaklinks=false,pdfborder={0 0 1},backref=false,colorlinks=false]
 {hyperref}

\makeatletter

\pdfpageheight\paperheight
\pdfpagewidth\paperwidth

\providecommand{\tabularnewline}{\\}

\@ifundefined{textcolor}{}
{%
 \definecolor{BLACK}{gray}{0}
 \definecolor{WHITE}{gray}{1}
 \definecolor{RED}{rgb}{1,0,0}
 \definecolor{GREEN}{rgb}{0,1,0}
 \definecolor{BLUE}{rgb}{0,0,1}
 \definecolor{CYAN}{cmyk}{1,0,0,0}
 \definecolor{MAGENTA}{cmyk}{0,1,0,0}
 \definecolor{YELLOW}{cmyk}{0,0,1,0}
}
\numberwithin{equation}{section}
\numberwithin{figure}{section}

\usepackage{babel}
\usepackage{calrsfs}
\usepackage{hyperref}
\hypersetup{
    colorlinks=true,
    linkcolor=red,
    citecolor=blue,
    filecolor=magenta,      
    urlcolor=cyan,
}
\makeatother

\begin{document}

\title{Feasible platform to study negative temperatures }

\author{R. J. de Assis}

\address{Instituto de Física, Universidade Federal de Goiás, 74.001-970, Goiânia
- GO, Brazil}

\author{C. J. Villas-Boas}

\address{Departamento de Física, Universidade Federal de São Carlos, 13565-905,
São Carlos, São Paulo, Brazil}

\author{N. G. de Almeida}

\address{Instituto de Física, Universidade Federal de Goiás, 74.001-970, Goiânia
- GO, Brazil}

\begin{abstract}
We afford an experimentally feasible platform to study Boltzmann negative
temperatures. Our proposal takes advantage of well-known techniques
of engineering Hamiltonian to achieve steady states with highly controllable
population inversion. Our model is completely general and can be applied
in a number of contexts, such as trapped ions, cavity-QED, quantum
dot coupled to optical cavities, circuit-QED, and so on. To exemplify,
we use Hamiltonian models currently used in optical cavities and trapped
ion domain, where the level of precision achieved the control of the
freedom degrees of a single atom inside a cavity/trapped ion. We show
several interesting effects such as absence of thermalization between
systems with inverted population and cooling by heating in these unconventional
systems.
\end{abstract}
\maketitle

\section{Introduction}

\textit{\emph{Negative temperature $T$ is associated with systems
with a bounded energy spectrum for which $1/T=(\partial S/\partial E)_{V}<0$,
where $S$ is the Boltzmann entropy. In general, negative temperatures
can be observed in the population-inverted states, which can be achieved
by injecting energy into these systems in the right way. The concept
of negative temperature was proposed in the early 1950s \cite{Purcell51,Ramsey56},
and since then has been discussed in a number of papers \cite{Tykodi75,Tremblay76,Tykodi76,Danielian76,Landsberg77,Tykodi78,Berdichevsky91,Rapp10,Mosk05}.
Very recently, negative temperature was experimentally observed using
ultracold quantum gases \cite{Braun13}. The experiment in \cite{Braun13}
triggered a vivid debate regarding the existence of negative absolute
temperatures, with the opponents claiming that negative absolute temperature
should not exist if the \textquotedbl{}right\textquotedbl{} (Gibbs)
entropy is used \cite{Sokolov14,Dunkel14,Hilbert14,Campisi15} while
those in favor of Boltzmann entropy claiming that negative temperatures
not only exist as it is necessary to explain systems presenting population
inversion \cite{Frenkel15,Anghel16,Vilar14,Buonsante16}. Although
this debate is far from a consensus, it seems to indicate that while
Boltzmann's entropy is adequate to describe the canonical ensemble
in the thermodynamical limit, Gibbs's entropy should be used to describe
the microcanonical ensemble \cite{Hilbert14,Campisi15} and, at least
tentatively, systems with finite dimension \cite{Dunkel14}. Anyway,
it is not our aim to enter in the debate on what is the right absolute
temperature, or, equivalently, the right entropy. Rather, here we
take a pragmatic approach of considering Boltzmann entropy, which
leads to temperatures that can be negative for systems with inverted
population.}}

Taking advantage of well-known techniques of Hamiltonian engineering
\cite{Prado06,Serra05}, in this paper we afford a platform experimentally
feasible in several contexts, as for example quantum circuits, cavity
QED and trapped ions. Particularly, here we explore a model in the
context of optical cavities and trapped ions, in which the extraordinary
level of precision reaches the control of individual atoms, in order
to engineer systems displaying negative Boltzmann temperature. To
our aim, we work with an ion trapped in a harmonic potential pumped
by a laser field to build steady states of the excited ionic levels,
therefore with inverted population. Similar inverted population can
also be achieved with single trapped atom inside an optical cavity,
by properly enginnering the atom-cavity mode interaction with the
help of external laser fields as done in \cite{Prado2009}. Within
our model, we are able to provide the range of values that negative
temperatures can be observed. After that, we study thermalization
of two systems when one of them, displaying negative temperatures,
is put in contact with a second one. We show that steady states of
systems involving negative temperatures can show very surprising behaviors,
as for example absence of thermalization with other systems.

\section{Model}

To our purpose, we make use of the so-called generalized anti-Jaynes-Cummings
model (AJCM) \cite{Leibfried03}, which, can be derived as follows.
Consider a trapped two-level ion whose transition frequency between
its excited and ground states is $\omega_{0}$ and the trap frequency
is $\nu$. The quantum of vibrational energy of the center of mass
of the ion is described by $a^{\dagger}a$. In the Schrödinger picture,
the Hamiltonian that describes such a system is $H=H_{f}+H_{a}+H_{int}\left(t\right)$,
with $H_{f}=\nu a^{\dagger}a$, $H_{a}=\omega_{0}\sigma_{z}/2$
\begin{equation}
H_{int}\left(t\right)=\frac{\Omega}{2}\left[\sigma_{-}e^{i\left(k\hat{x}+\omega_{L}t\right)}+\sigma_{+}e^{-i\left(k\hat{x}+\omega_{L}t\right)}\right],\label{Hint}
\end{equation}
where $\hbar=1$ and the Rabi frequency (ion-laser coupling) $\Omega$
is much smaller than the bosonic ($\nu$) and atomic transition ($\omega_{0}$)
frequencies; $\sigma_{+}$($\sigma_{-}$) is the raising (lowering)
Pauli operator for a two-level ionic system, $\sigma_{z}=\sigma_{+}\sigma_{-}-\sigma_{-}\sigma_{+}$,
$a$ ($a^{\dagger}$) is the annihilation (creation) operator in the
Fock space for the bosonic mode, and $\omega_{L}$ is the frequency
of the driving laser. Here $k\hat{x}=\eta_{L}(a+a^{\dagger})$, and
$\eta_{L}=k/\sqrt{2m\nu}$ is the Lamb-Dicke parameter \cite{Leibfried03}.
Working in the limit $\eta_{L}\ll1$ and applying the rotating-wave
approximation, the Hamiltonian $H_{int}\left(t\right)$ in the interaction
picture has the time-independent form 
\begin{equation}
H_{I}=g_{k}\left(\sigma_{-}a^{k}+\sigma_{+}a^{\dagger k}\right),\label{AJC}
\end{equation}
where, by adjusting the frequency $\omega_{L}$ such that $\delta=\omega_{L}-\omega_{0}=k\nu$
on resonance with the two-level ion, we can have, for instance, interactions
i) carrier ($k=0,\:g_{0}=\Omega/2$) ii) the first ($k=1,\:g_{1}=i\Omega\eta_{L}/2$)
and iii) second ($k=2,\:g_{2}=-i\eta_{L}^{2}\Omega/4$) blue sideband,
and so on \cite{Rossato12,Ruynet95}. Note that, as we are working
in the $\eta\ll1$ limit, the nonlinear dependence of the coupling
with $a^{\dagger}a$ can be neglected \cite{Ruynet95}. This Hamiltonian
experimentally model a number of systems, as for example trapped ions
\cite{Leibfried03,Poyatos96}, a two-level (TL) atom pumped by a classical
electromagnetic field (EM) and interacting with a quantized mode of
a EM \cite{Rosado15}, a coupling between spin and nanomechanical
resonator \cite{xue07} or either a TL neutral atom in a dipole trap
or a TL ion in a harmonic trap \cite{Rempe00,Kimble03,Blatt09}. The
dynamics of this model for weak system-reservoir coupling can be described
by the master equation formalism, which for the Hamiltonian (Eq. \eqref{AJC})
reads
\begin{align}
\frac{\partial\rho}{\partial t} & =-i\left[H_{I},\rho\right]+\kappa\left(n_{f}+1\right)\mathcal{D}[a]\rho+\kappa n_{f}\mathcal{D}[a^{\dagger}]\rho\nonumber \\
 & +\gamma\left(n_{a}+1\right)\mathcal{D}[\sigma_{-}]\rho+\gamma n_{a}\mathcal{D}[\sigma_{+}]\rho\label{Master}
\end{align}
where $\kappa$ and $\gamma$ are the spontaneous emission rates for
the vibrational mode and atom, respectively, $n_{f}$ ($n_{a}$) is
the average photon number for the vibrational mode (atom) reservoir,
and $D[A]\rho\equiv2A\rho A^{\dagger}-A^{\dagger}A\rho-\rho A^{\dagger}A$.

Next, we solve numerically Eq. \eqref{Master} to obtain the steady
state of the system at $t\rightarrow\infty$ by imposing $\partial\rho/\partial t=0$
and thus to calculate the corresponding thermodynamic properties.
We note that for solving numerically this system we must truncate
the infinite Fock basis of the bosonic fields somewhere, which depend
on the mean number of excitations in the bosonic field (vibrational
mode). To proceed a numerical study of Eq. \eqref{Master} \cite{Tam99},
we assume both reservoirs with the same average number of thermal
photons, i.e., $n_{f}=n_{a}=n$. Also, as we are particularly interested
on the TL system, whose asymptotic behavior can present inverted population,
we trace over the bosonic field variables. Next, we use the master
Eq. \eqref{Master} to investigate negative temperatures for a range
of parameters represented by the cooperativity $C_{k}=g_{k}^{2}/\gamma\kappa$.
The cooperativity is the experimentally relevant parameter and depends
on the physical context. For instance, in the context of optical cavities
it was experimentally realized up to $C\sim35$ \cite{Rempe00}.

It is important to mention some differences and similarities between
our model and the system used in \cite{Braun13}. As in \cite{Braun13},
we use external fields to engineer an effective Hamiltonian which
drives the system to a stationary regime with population inversion.
In the experiment with cold atoms in optical latices described in
\cite{Braun13}, via Feshbach's resonance, an effective interaction
is engineered to prepare a steady state with negative temperature,
that is, starting from a state without population inversion, the system
reaches a state with population inversion for sufficiently long times.
However, it is important to note that the relevant dissipative mechanics
considered in \cite{Braun13} are sufficiently weak such that they
can be disregarded in that experiment, allowing to treat the atomic
ensemble as an isolated system. In our model, the dissipation of the
atom is taken into account and it is responsible for destroying the
state with population inversion. However, the engineered effective
interaction with the bosonic mode plus its dissipation result in an
effective engineered reservoir which leads the system to a state with
inverted population. Thus, as we take into account the ionic dissipation
in our model, here we have a competition between the action of the
natural dissipation of the ion (spontaneous emission) and the action
of the engineered interaction.

\section{Producing steady states with negative temperatures}

\begin{figure}
\begin{centering}
\begin{tabular}{cc}
\includegraphics[bb=20bp 0bp 570bp 395bp,scale=0.21]{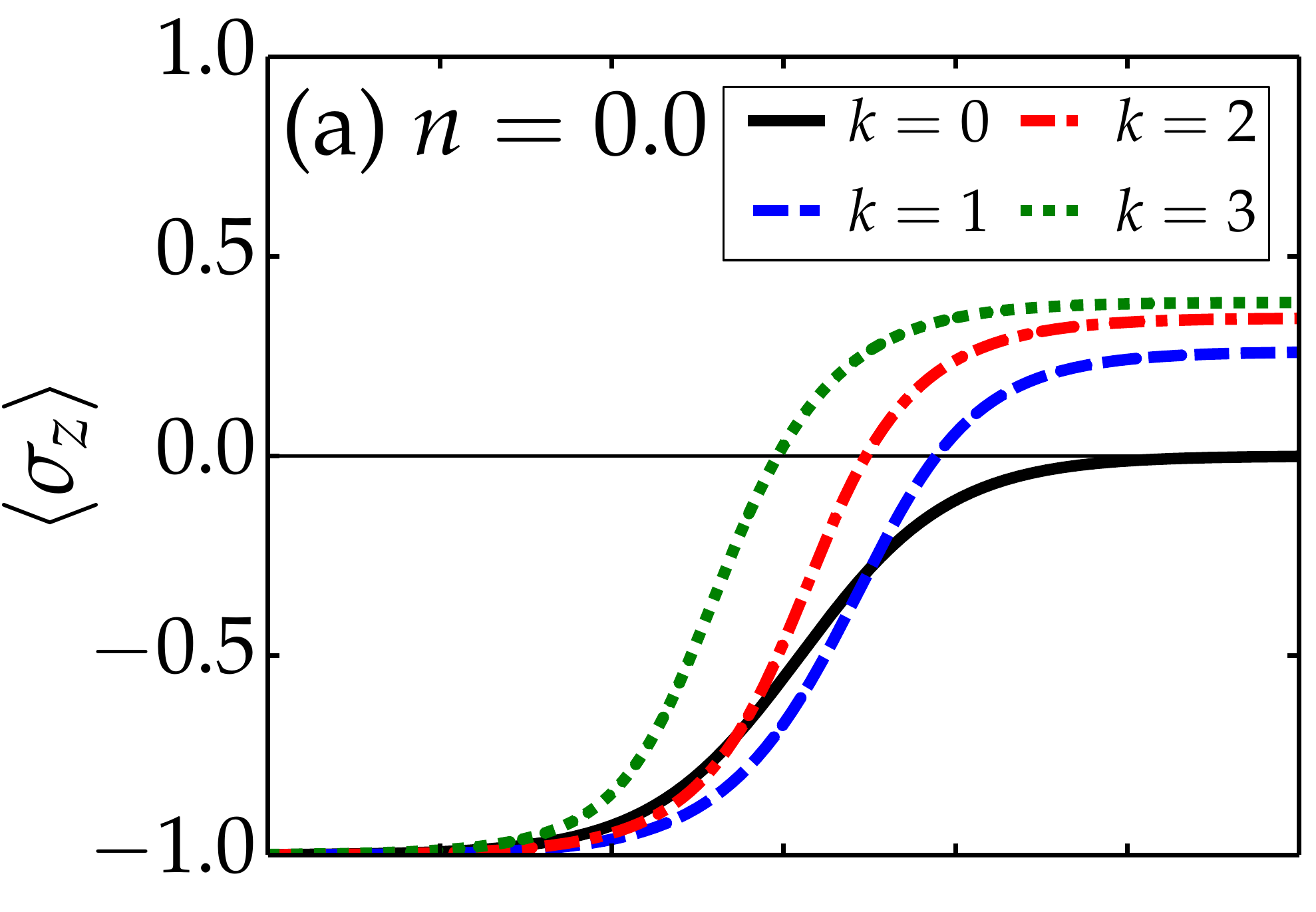} & \includegraphics[bb=50bp 0bp 560bp 395bp,scale=0.21]{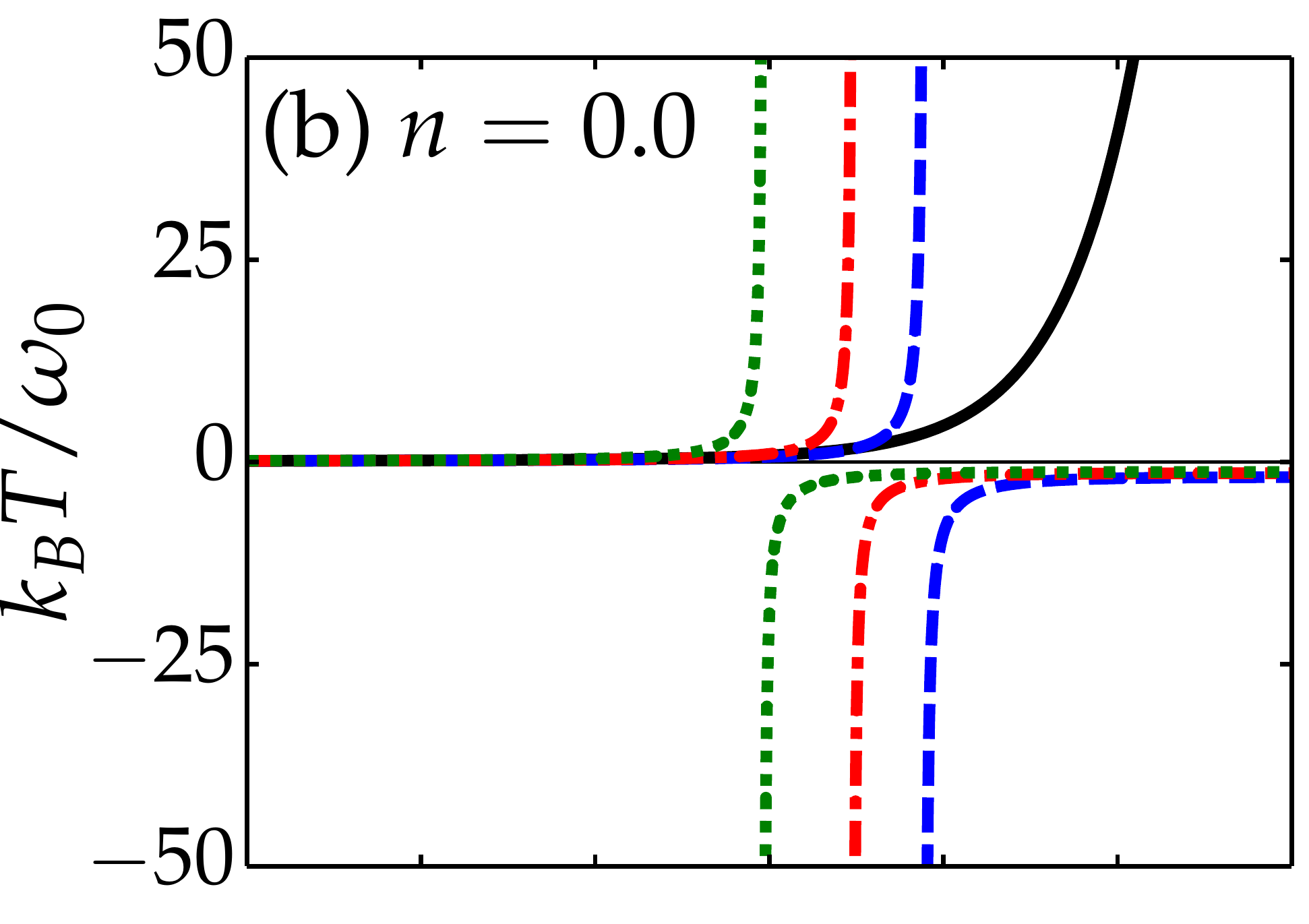}\tabularnewline
\includegraphics[bb=20bp 0bp 570bp 395bp,scale=0.21]{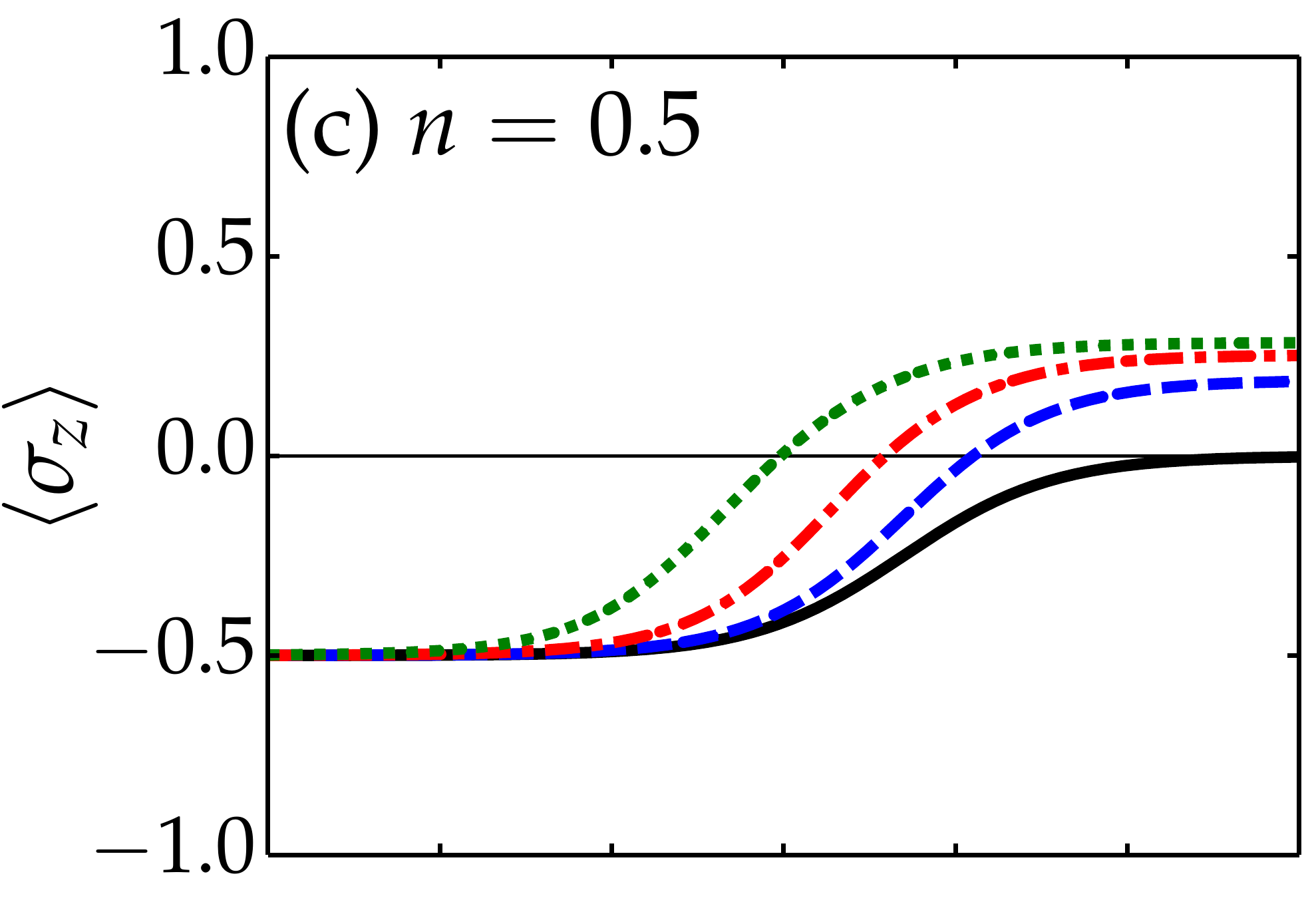} & \includegraphics[bb=50bp 0bp 560bp 395bp,scale=0.21]{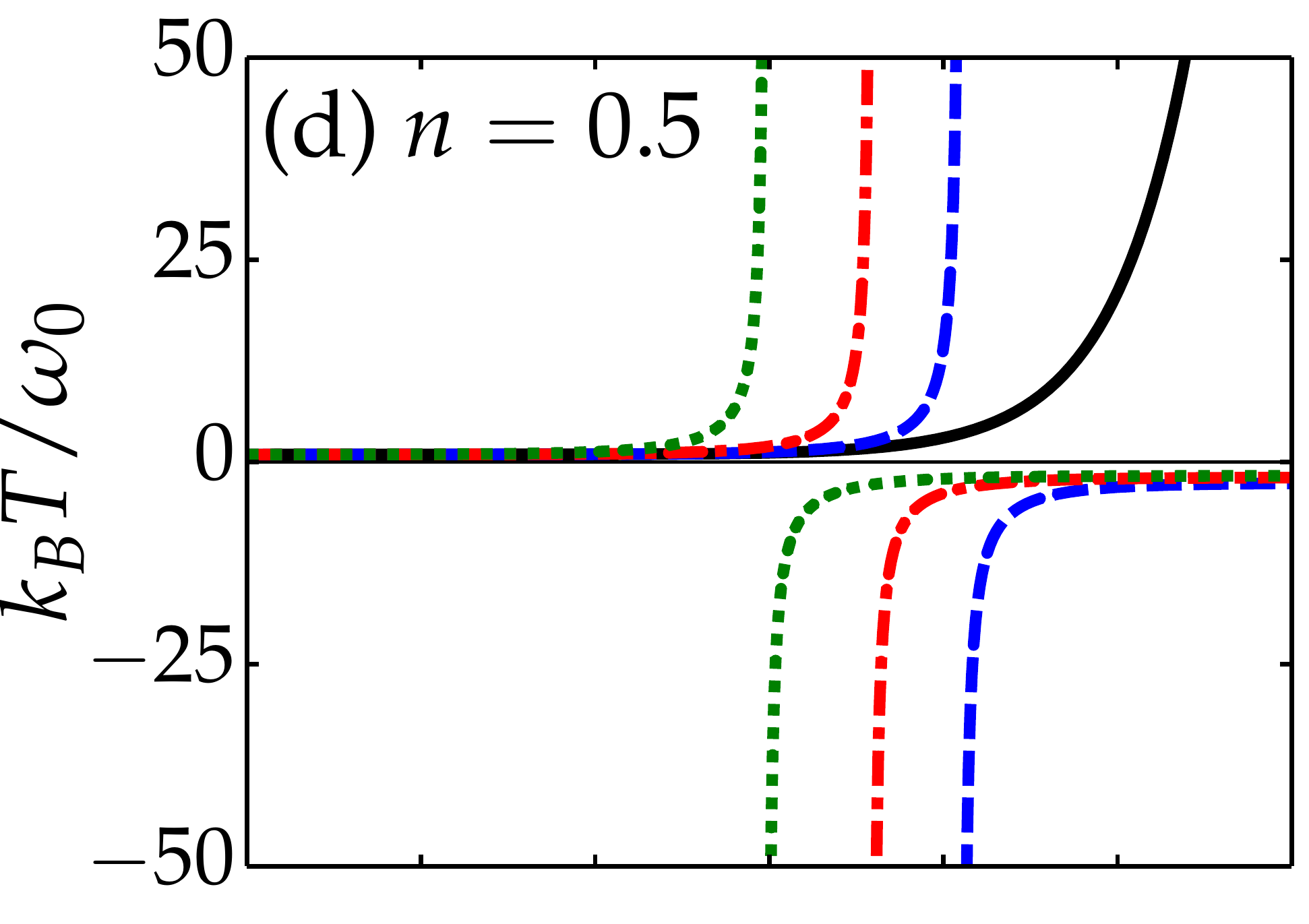}\tabularnewline
\includegraphics[bb=5bp 0bp 586bp 463bp,scale=0.21]{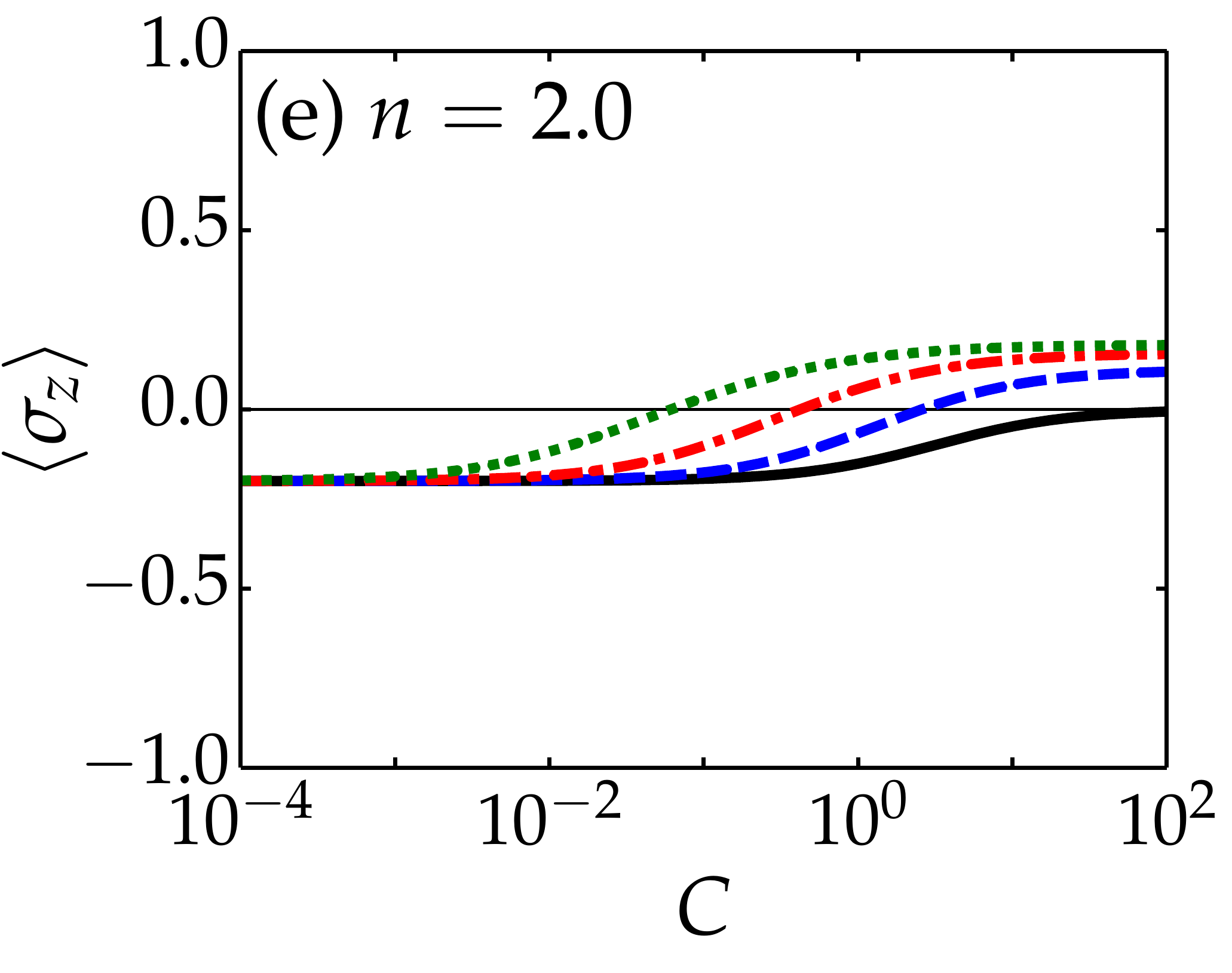} & \includegraphics[bb=20bp 0bp 587bp 448bp,scale=0.21]{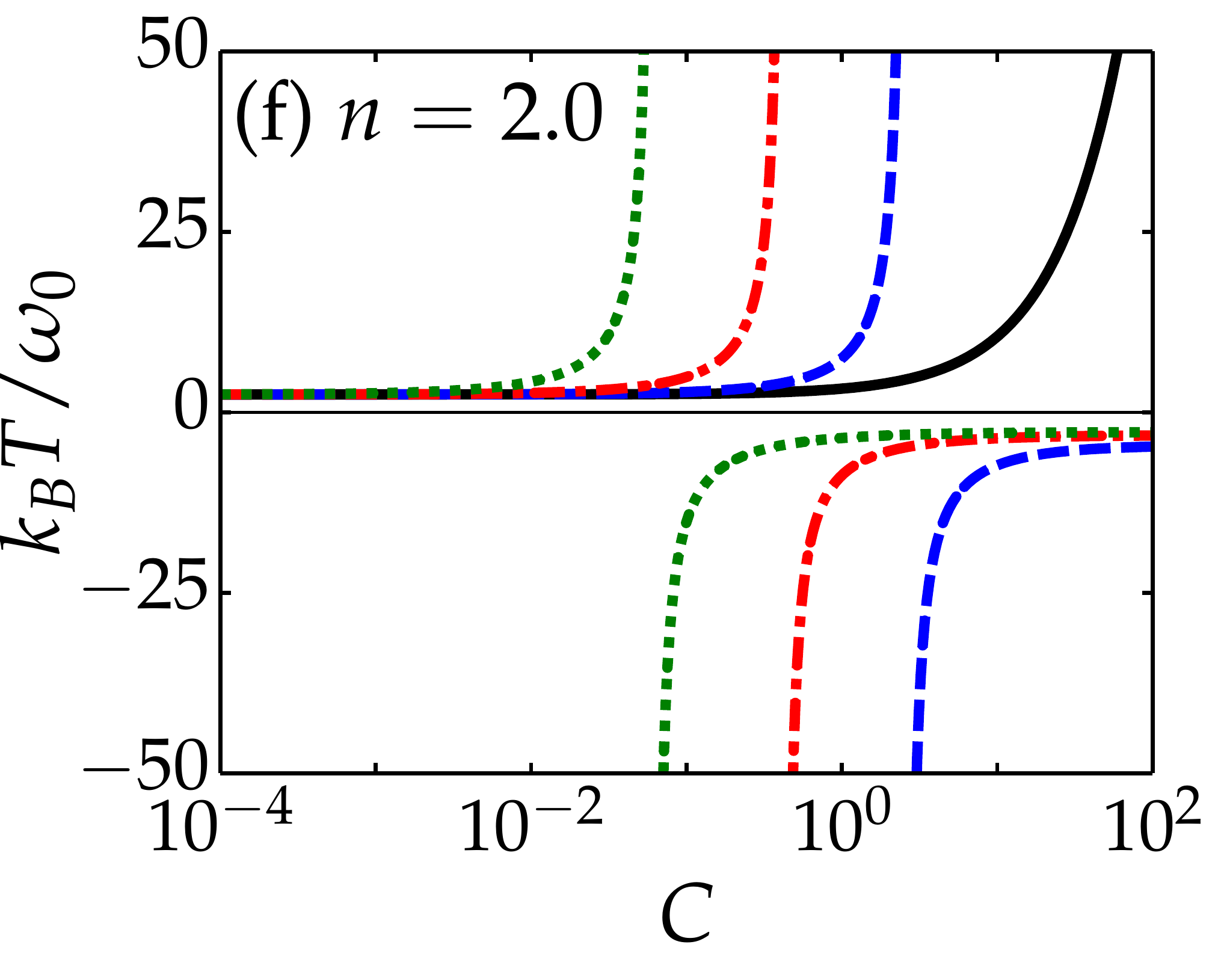}\tabularnewline
\end{tabular}
\par\end{centering}
\centering{}\caption{\label{Fig1}Atom scaled average energy $\left\langle \sigma_{z}\right\rangle $
\emph{versus} cooperativity $C$, Figs. (a,c,e), and its scaled effective
temperature $k_{B}T/\omega_{0}$ \emph{versus} cooperativity $C$,
Figs. (b,d,f). The environment thermal photon averages are $n=0.0$,
$0.5$ and $2.0$, respectively. The Hamiltonian models are $k=0$
(solid black line), $k=1$, (dashed blue line), $k=2$ (dot-dashed
red line) and $k=3$ (dotted green line). Inverted population occurs
for all $k$ except for $k=0$. Here we used $\gamma=\kappa$.}
\end{figure}

Let us begin by investigating how the energy and the corresponding
effective temperature of the TL system varies with the cooperativity.
As said before, we are interested in the steady state, such that we
consider the asymptotic limit $t\rightarrow\infty$ by imposing $\partial\rho/\partial t=0$
for $k=0,1,2,3$. In Figs. \ref{Fig1}(a-f), where the cooperativity
parameter is displayed in logarithmic scale for clarity, we have considered
the environments with average thermal photons $n=0$, $0.5$, and
$2.0$. From Figs. \ref{Fig1}(a,c,e) we see that, except for $k=0$
(solid black line), all steady states end with inverted population
$\left\langle \sigma_{z}\right\rangle >0$, thus leading to negative
temperatures, as shown in Figs. \ref{Fig1}(b,d,f), where the rescaled
temperature $k_{B}T/\omega_{0}$ \emph{versus} cooperativity is shown.
Here $T$ was obtained through\textbf{\textit{\emph{ $T=1/k_{B}\left(dS_{a}/d\left\langle H_{a}\right\rangle \right)$,
}}}\textit{\emph{where $S_{a}$ is the von Neumann entropy to the
ion. }}Note that the role of nonlinearity is to populate the excited
state, thus enhancing the inverted population: the greater $k$, the
greater the positive average $\left\langle \sigma_{z}\right\rangle $,
which is also reflected in the negative temperature. It is interesting
to note that inverted population can be obtained with small values
of cooperativity simply increasing the nonlinearity of the AJCM. Also,
note that the bosonic and atomic environments slightly suppress the
inversion of population, Figs. \ref{Fig1}(a,c,e), thus avoiding to
reach hotter negative temperatures for the TL system. From Figs. \ref{Fig1}(b,d,f)
we can see that negative temperatures approach to $0K$ to all $k=1,2,3$
and for sufficiently high values of the cooperativity. A numerical
inspection shows us that negative temperatures \textit{only} occur
for $C\gtrsim0.65$.

\section{Negative temperatures and coupled systems}

Here we show how our platform provides an excellent tool for studying
thermalization of two coupled systems and how to control the heating
or cooling of both systems through cooperativity. Since systems presenting
inverted population is hotter than other systems with positive temperatures
\cite{Note}, it is interesting to understand what happens when these
\emph{hotter than hot} \cite{Note2} systems are coupled to each other.
Let us consider a two-level atom (A) coupled to another one (B) through
a simple exchange interaction $H_{N}=\lambda(\sigma_{+}^{A}\sigma_{-}^{B}+\sigma_{-}^{A}\sigma_{+}^{B})$,
which is the usual interaction resulting, for example, of a collision
process \cite{Zheng00}. The atom A is coupled to a quantum bosonic
mode through the effective AJCM Eq. \eqref{AJC}\textbf{ }and interacts
via $H_{N}$ with atom B. For simplicity, we assume the vibrational
mode, the atom A, and the atom B decaying at the same rate ($\gamma_{A}=\gamma_{B}=\kappa=\gamma$).

To study thermalization in these systems, we trace out the degrees
of freedom of the bosonic mode for each model $k=0,1,2$ and $3$
separately. The case $k=0$, although presenting no population inversion,
is considered here only for the effect of comparing thermalization
to both atoms A and B. 

\begin{figure}[ptbh]
\centering{}%
\begin{tabular}{cc}
\includegraphics[bb=20bp 0bp 570bp 395bp,scale=0.21]{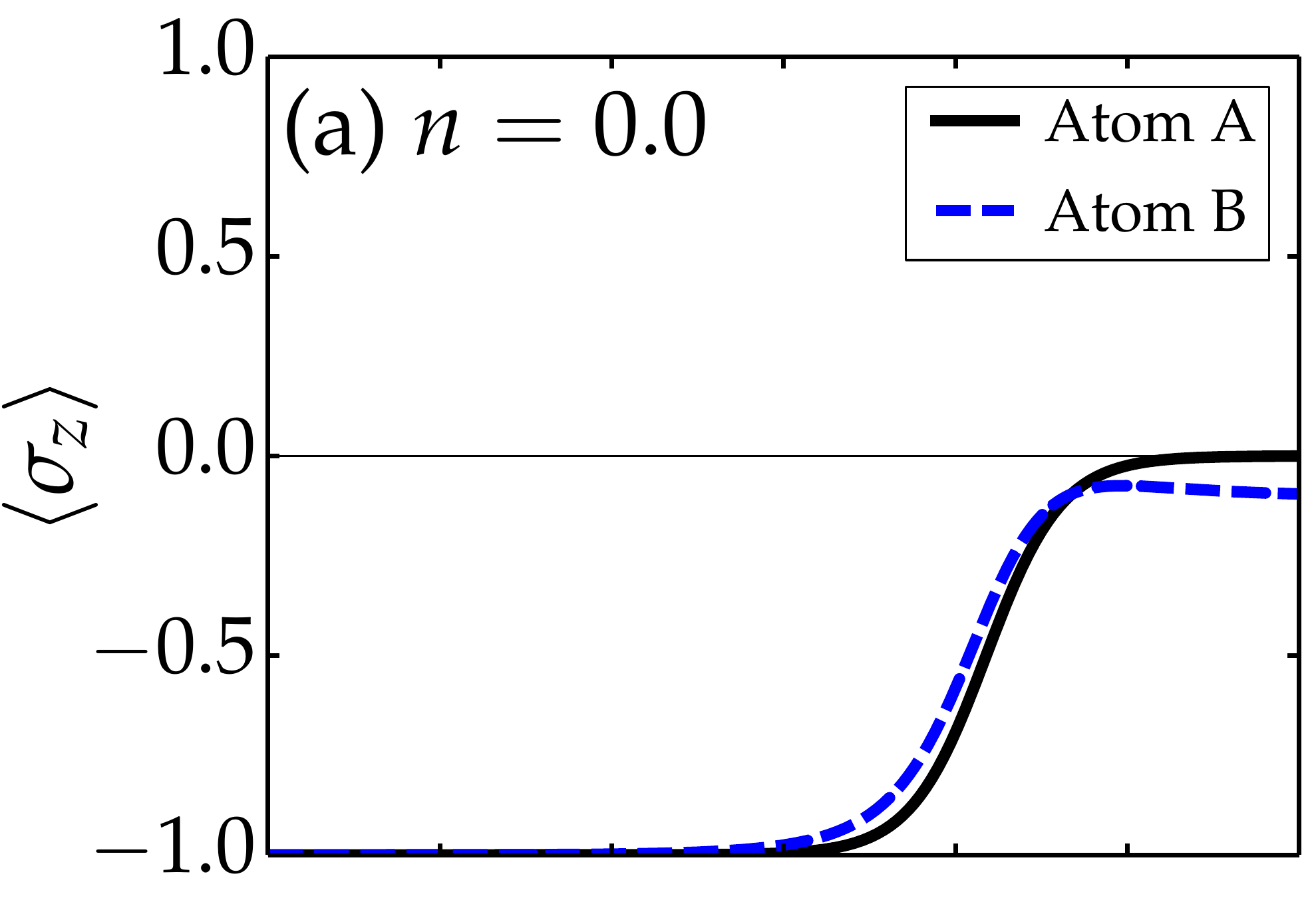} & \includegraphics[bb=50bp 0bp 559bp 395bp,scale=0.21]{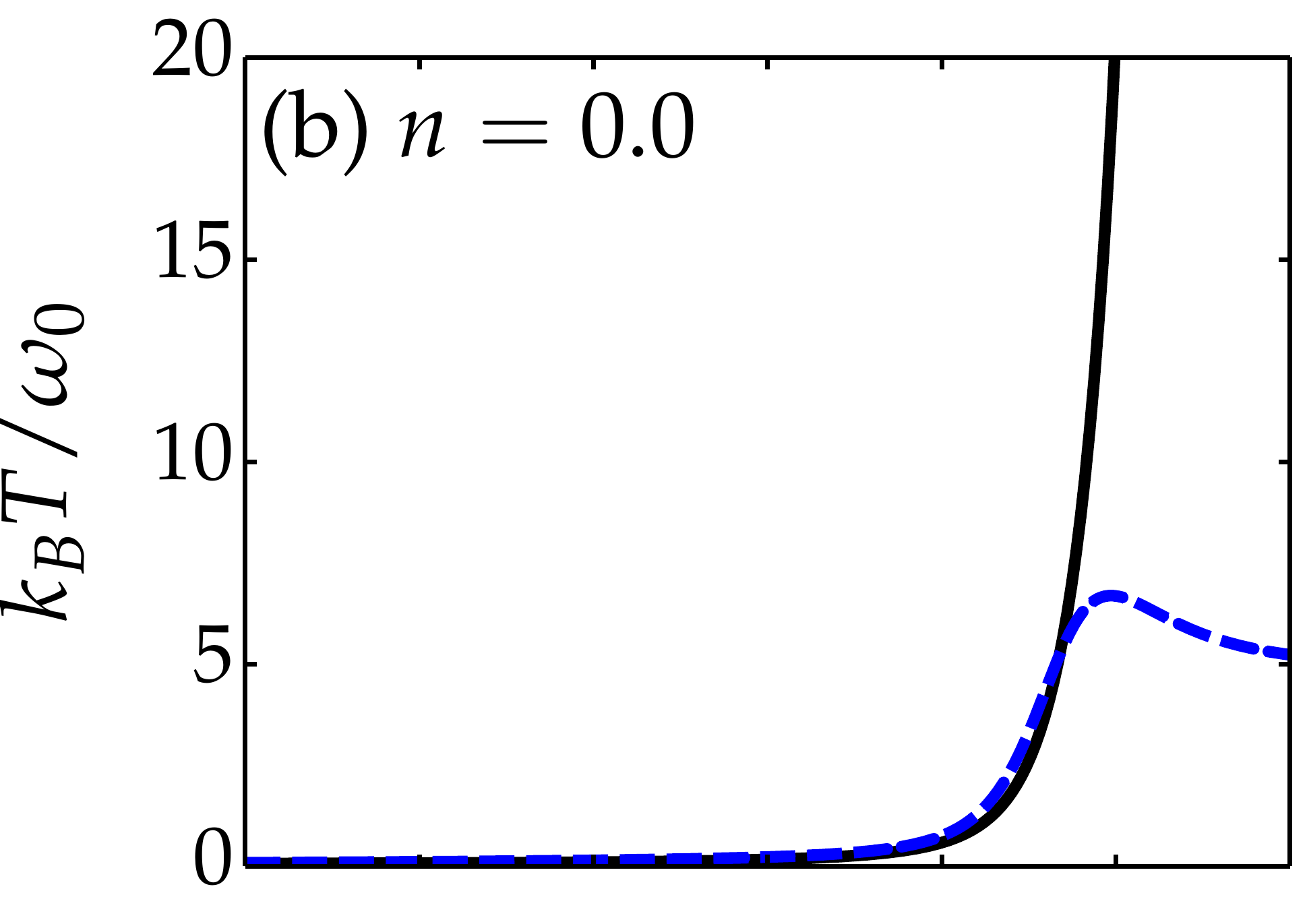}\tabularnewline
\includegraphics[bb=20bp 0bp 570bp 395bp,scale=0.21]{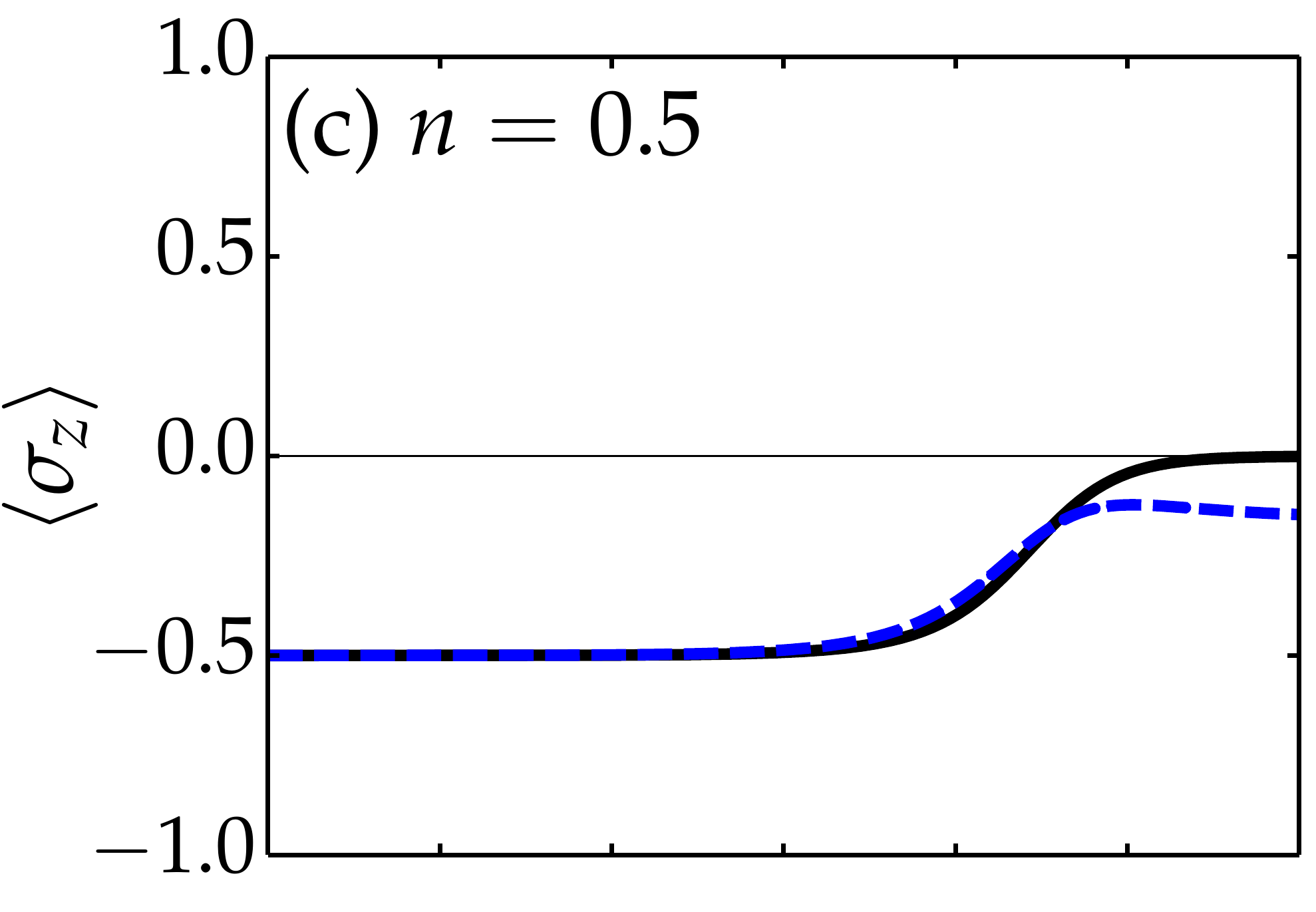} & \includegraphics[bb=50bp 0bp 559bp 395bp,scale=0.21]{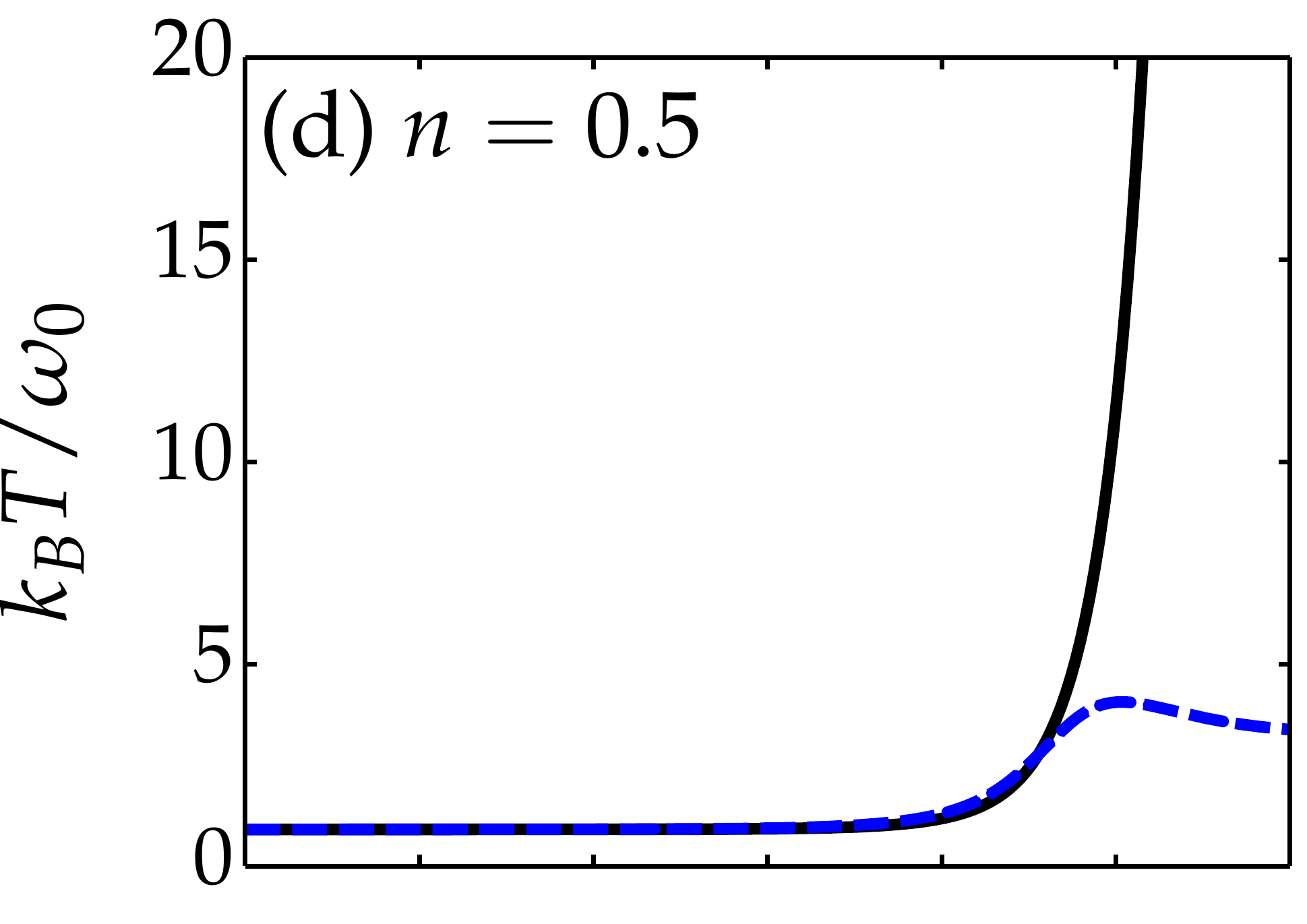}\tabularnewline
\includegraphics[bb=5bp 0bp 586bp 463bp,scale=0.21]{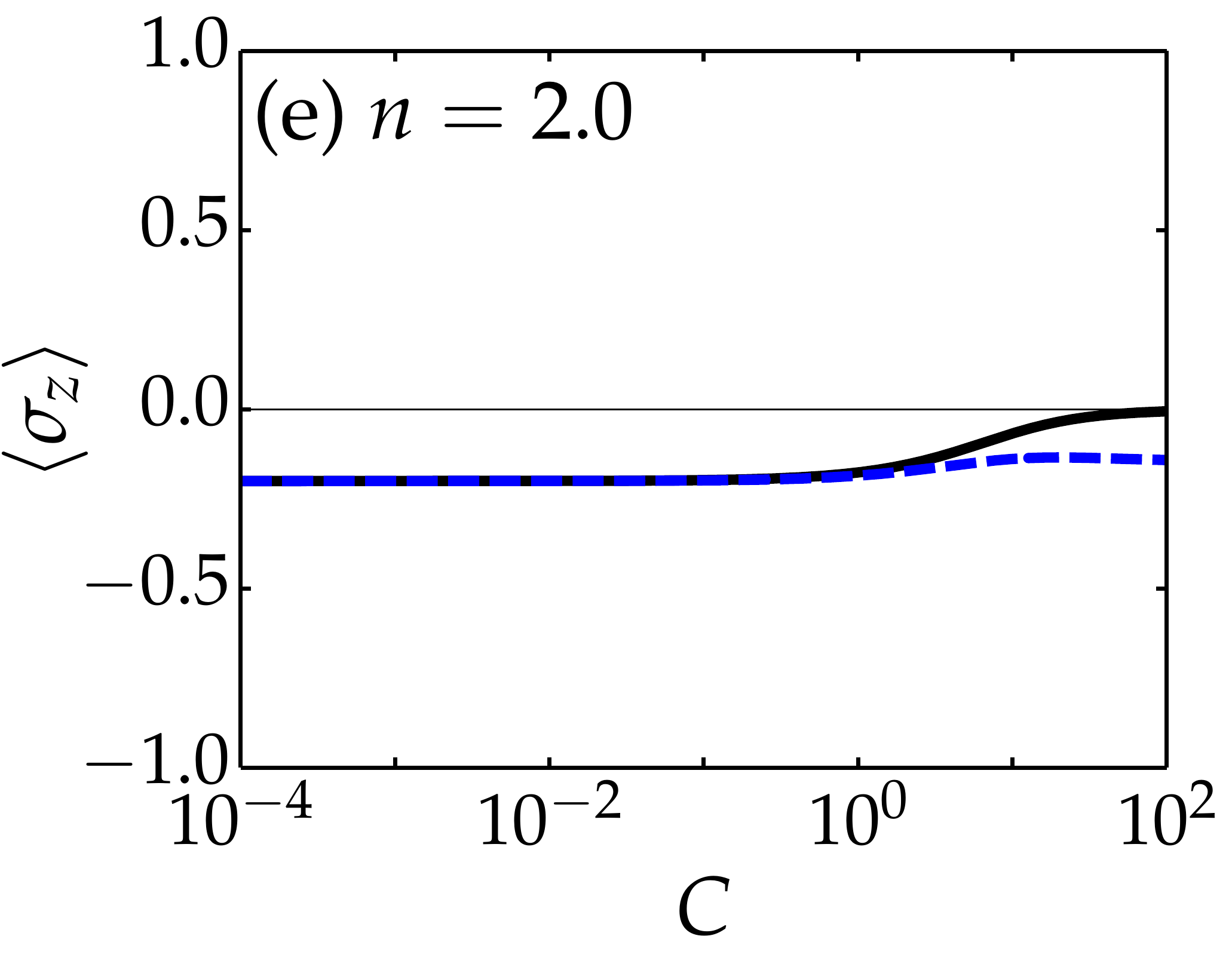} & \includegraphics[bb=20bp 0bp 586bp 463bp,scale=0.21]{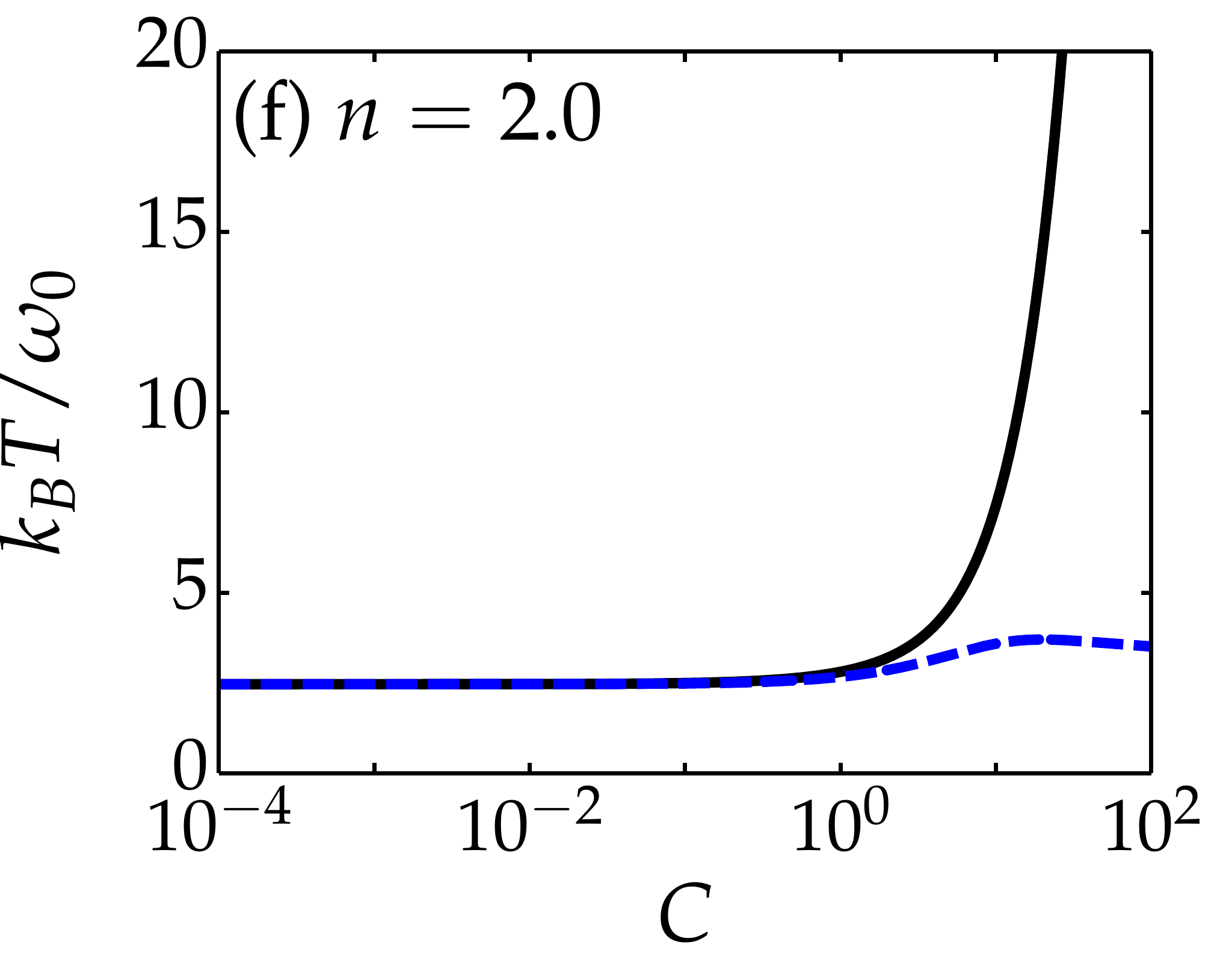}\tabularnewline
\end{tabular}\caption{\label{Fig2}Atom scaled average energy $\left\langle \sigma_{z}\right\rangle $
\emph{versus} cooperativity, Figs. (a,c,e), and its scaled effective
temperature $k_{B}T/\omega_{0}$ \emph{versus} cooperativity, Figs.
(b,d,f), for $k=0$. The environment thermal photon averages are $n=0.0$,
$0.5$, and $2.0$, respectively. Curves for atom A (B) is indicated
by solid black (blue dashed) line. Here we used $\lambda=3\gamma$
and $\gamma=\gamma_{A}=\gamma_{B}=\kappa$. }
\end{figure}

In Figs. \ref{Fig2} we show the average energy for each atom \emph{versus}
cooperativity, Figs. \ref{Fig2}(a,c,e), and the scaled effective
temperature (for each atom) \emph{versus} cooperativity, Figs. \ref{Fig2}(b,d,f),
for $k=0$ considering the environment thermal photon averages $n_{f}=n_{a}^{A}=n_{a}^{B}=n=0,0.5$
and $2.0$. As seen from these figures, the populations is not inverted.
Also, for $\lambda=3\gamma$, from a certain value of the cooperativity,
$C\gtrapprox8$, as the energy is not the same for both atoms, Figs.
\ref{Fig2}(a,c,e), the effective temperature, Fig. \ref{Fig2}(b,d,f),
also will be not the same, thus indicating that for $C\gtrapprox8$
there will be no thermalization. As indicated by our numerical simulation,
not shown, the greater $\lambda$, the greater the cooperativity required
for the atom-atom system to fail to present thermalization.\textbf{
}The effect of increasing the environment temperature is to produce
steady states with higher temperatures, without producing population
inversion. 

\begin{figure}[ptbh]
\centering{}%
\begin{tabular}{cc}
\includegraphics[bb=20bp 0bp 570bp 395bp,scale=0.21]{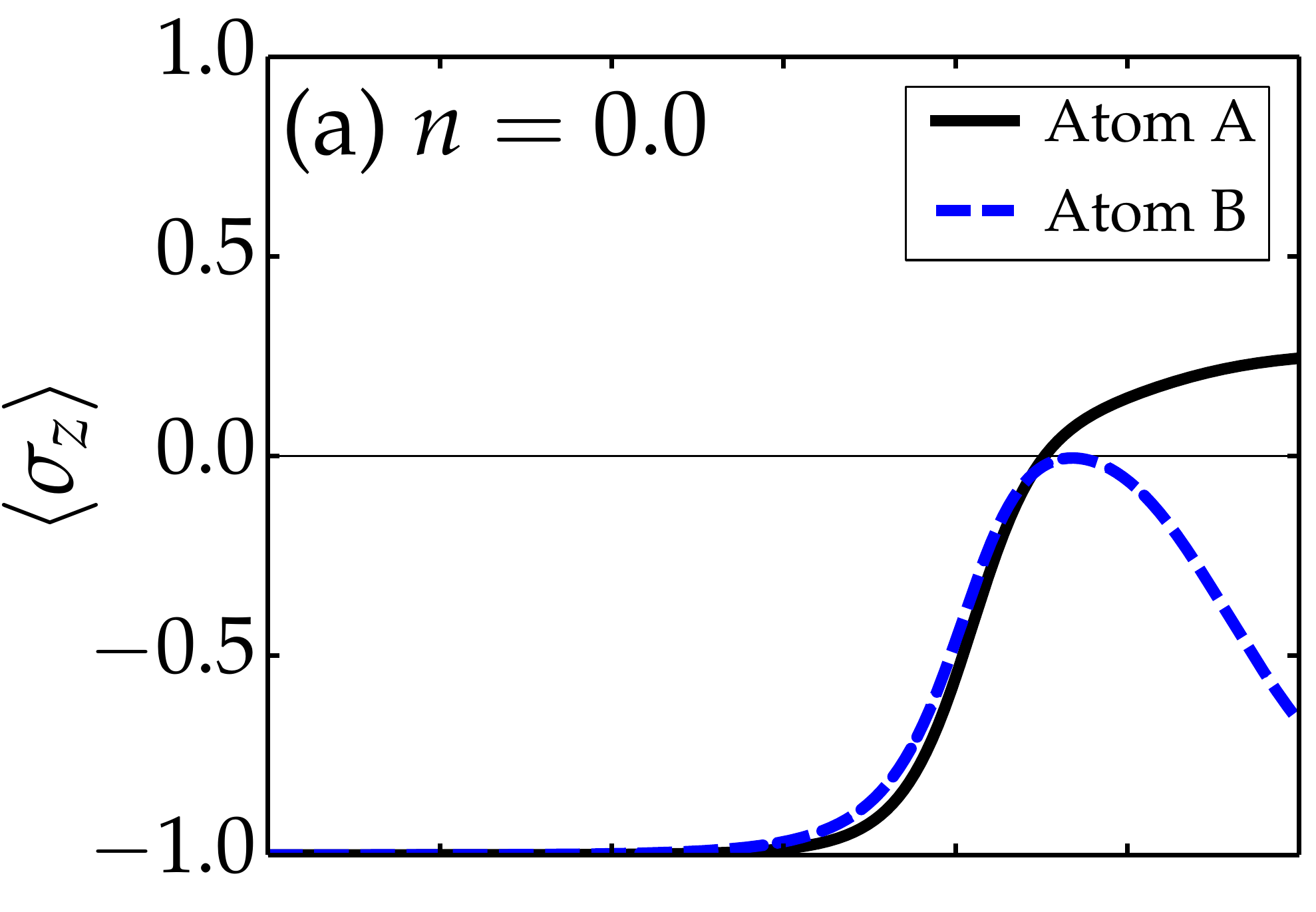} & \includegraphics[bb=50bp 0bp 559bp 395bp,scale=0.21]{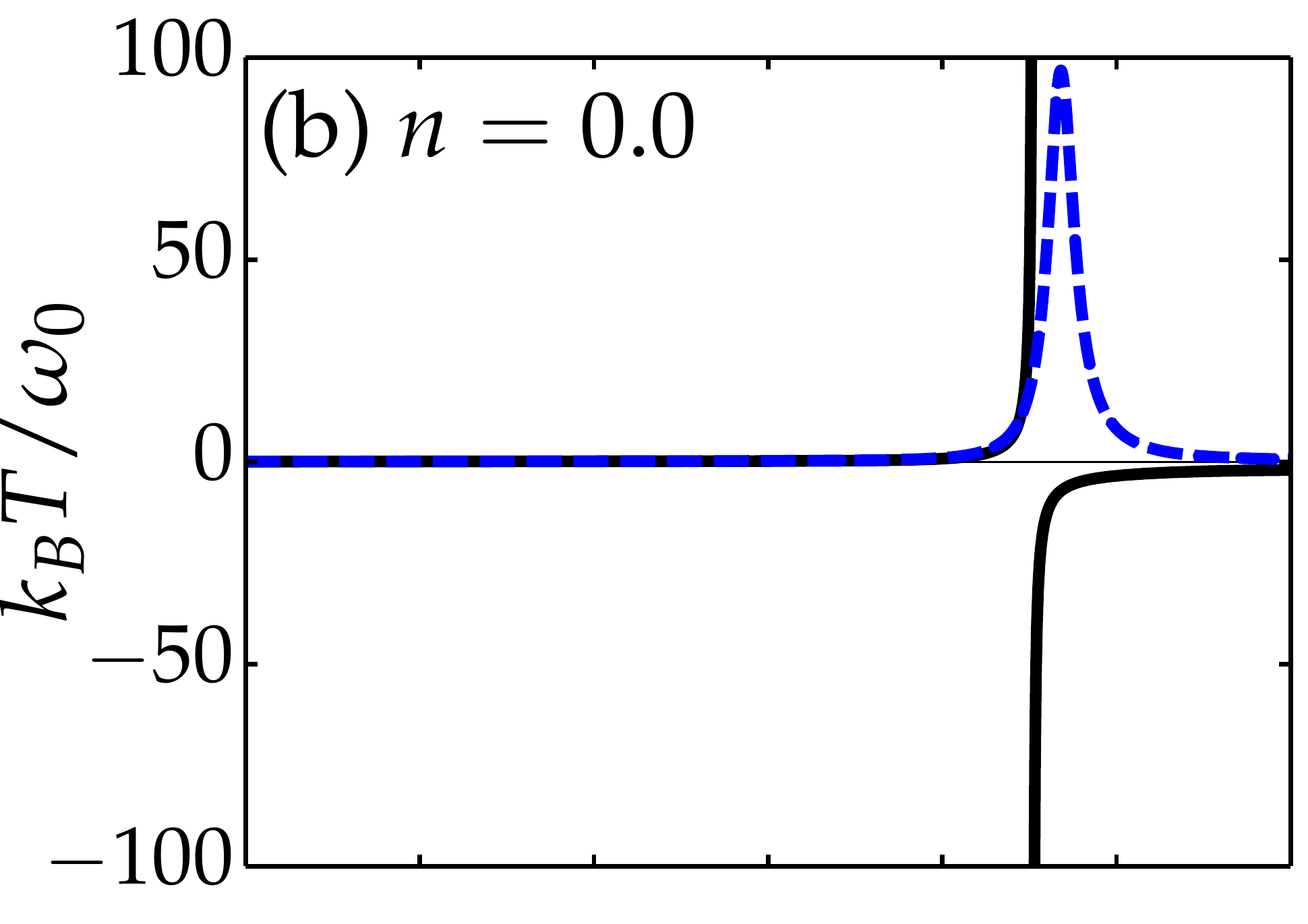}\tabularnewline
\includegraphics[bb=20bp 0bp 570bp 395bp,scale=0.21]{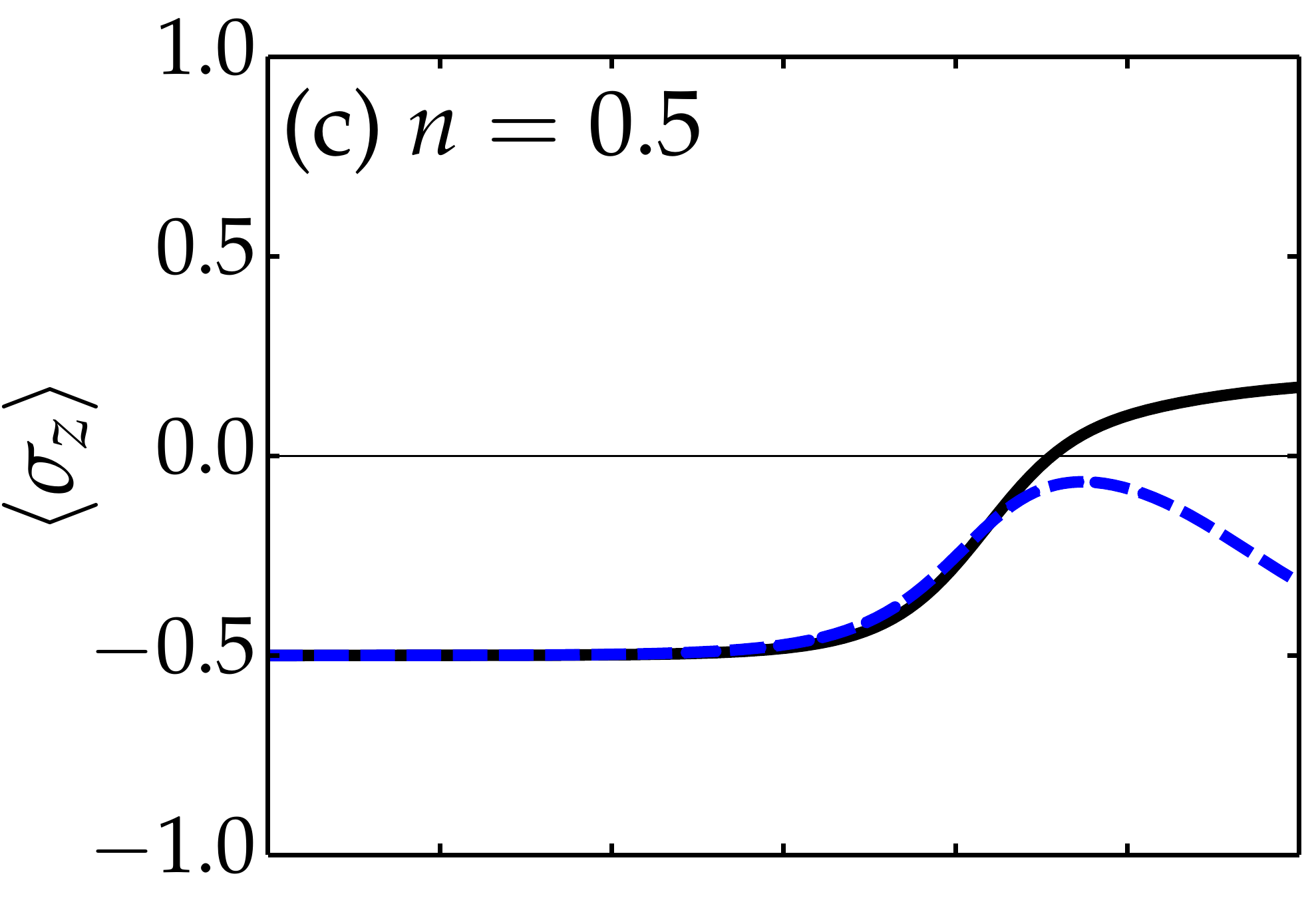} & \includegraphics[bb=50bp 0bp 560bp 395bp,scale=0.21]{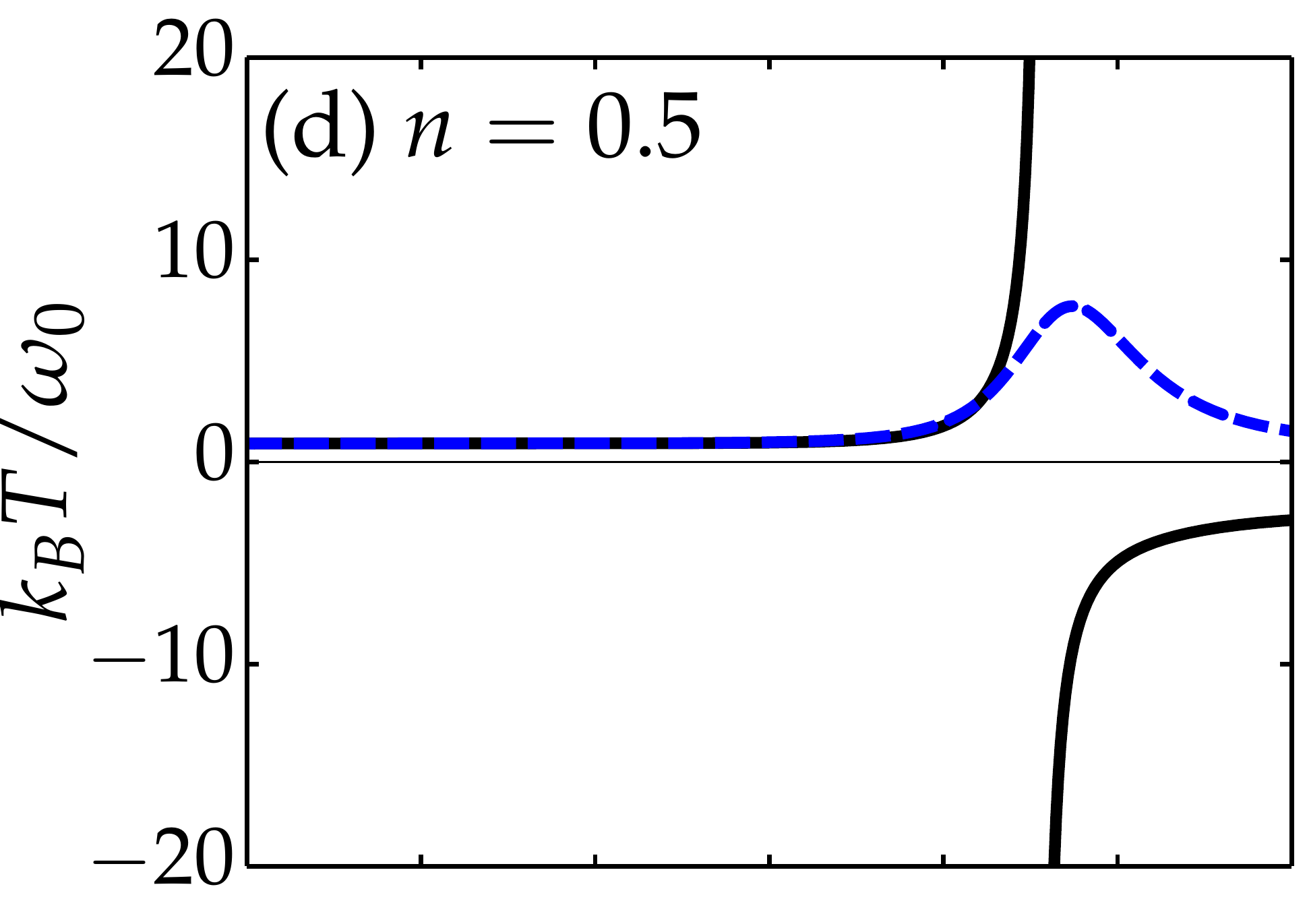}\tabularnewline
\includegraphics[bb=5bp 0bp 586bp 463bp,scale=0.21]{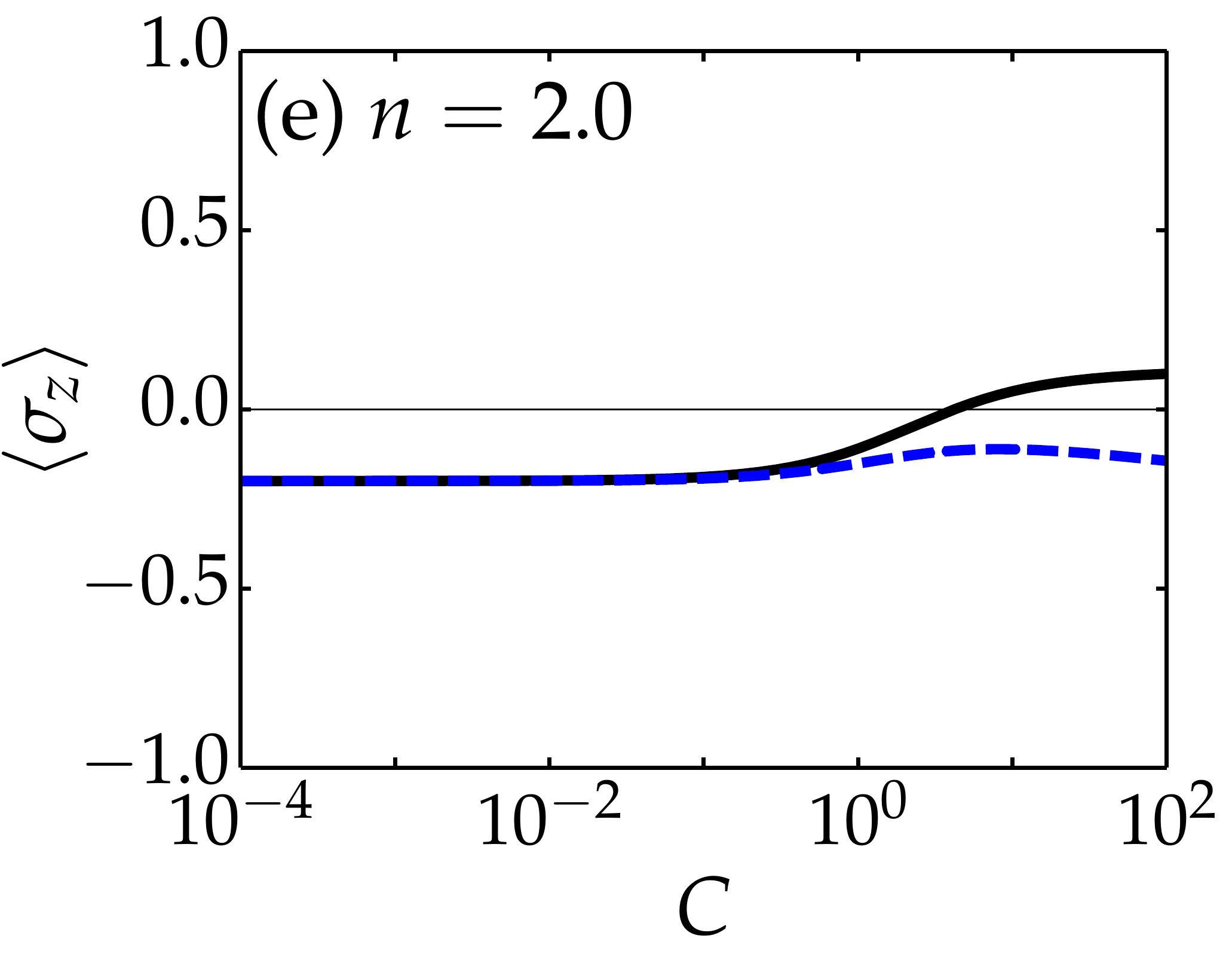} & \includegraphics[bb=20bp 0bp 586bp 463bp,scale=0.21]{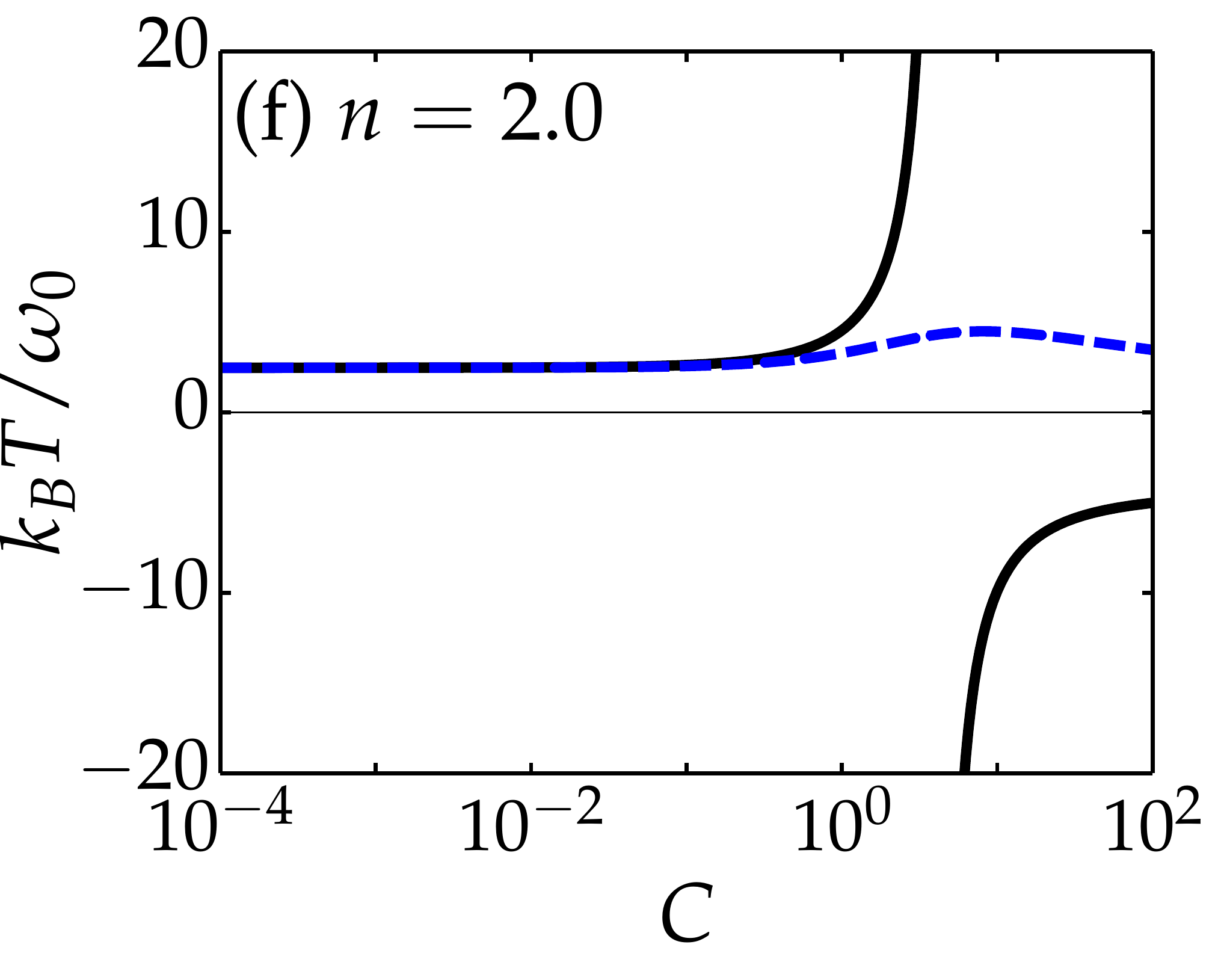}\tabularnewline
\end{tabular}\caption{\label{Fig3}Scaled average energy $\left\langle \sigma_{z}\right\rangle $
\emph{versus} cooperativity, Figs. (a,c,e), and scaled effective temperature
$k_{B}T/\omega_{0}$ \emph{versus} cooperativity, Figs. (b,d,f), for
$k=1$. The environment thermal photon averages are $n=0.0$, $0.5$,
and $2.0$, respectively. Curves for atom A (B) is indicated by solid
black (blue dashed) line. Here we used $\lambda=3\gamma$ and $\gamma=\gamma_{A}=\gamma_{B}=\kappa$.
A similar behavior can be seen for models $k=2,3$.}
\end{figure}

In Fig.\ref{Fig3} we show the average energy \emph{versus} cooperativity,
Figs. \ref{Fig3}(a,c,e), and the scaled effective temperature \emph{versus}
cooperativity, Figs. \ref{Fig3}(b,d,f), for $k=1$ considering the
environment thermal photon averages $n=0$, $0.5$, and $2.0$. The
behavior of these curves, although similar to those in Figs. \ref{Fig2}(a-f),
now show that atom A presents population inversion from $C\gtrsim3.5$
different from atom B which always has positive temperatures (for
$\lambda=3\gamma$). Also, for a certain value of $C$ the energies
of atoms A and B are not the same, thus indicating that there will
be no thermalization. Here, different from $k=0$, the effect of increasing
the environment temperature is to produce steady states with lower
negative temperatures. In other words, the positive reservoir tends
to lowering the negative temperature of atom A. Another surprising
effect is that temperature does not increase monotonically with cooperativity.
This effect can be better appreciated looking to the atom B curve
in Figs. \ref{Fig3}(b): by increasing the cooperativity, the temperature
of atom B, as given by the dashed blue line, first increase until
$C\thickapprox4.5$ attaining its maximum, and then decays to zero
for large values of the cooperativity.\textbf{ }

The models $k=2,3$ presents a similar pattern an will not be shown
here: atom B, which is coupled only to atom A, never inverts its population,
but cannot be heated arbitrarily, since its temperature reaches a
maximum. Otherwise, atom A, which is coupled both to atom B and to
a bosonic mode, is the one that has its population inverted, being
this population (and so its negative temperature) diminished when
the temperature of the environment reservoir is increased (from $n=0$
to $2.0$). Also, the atomic steady states do not thermalize neither
with each other nor with the environment.

\begin{figure}[ptbh]
\centering{}%
\begin{tabular}{cc}
\includegraphics[bb=20bp 0bp 570bp 395bp,scale=0.21]{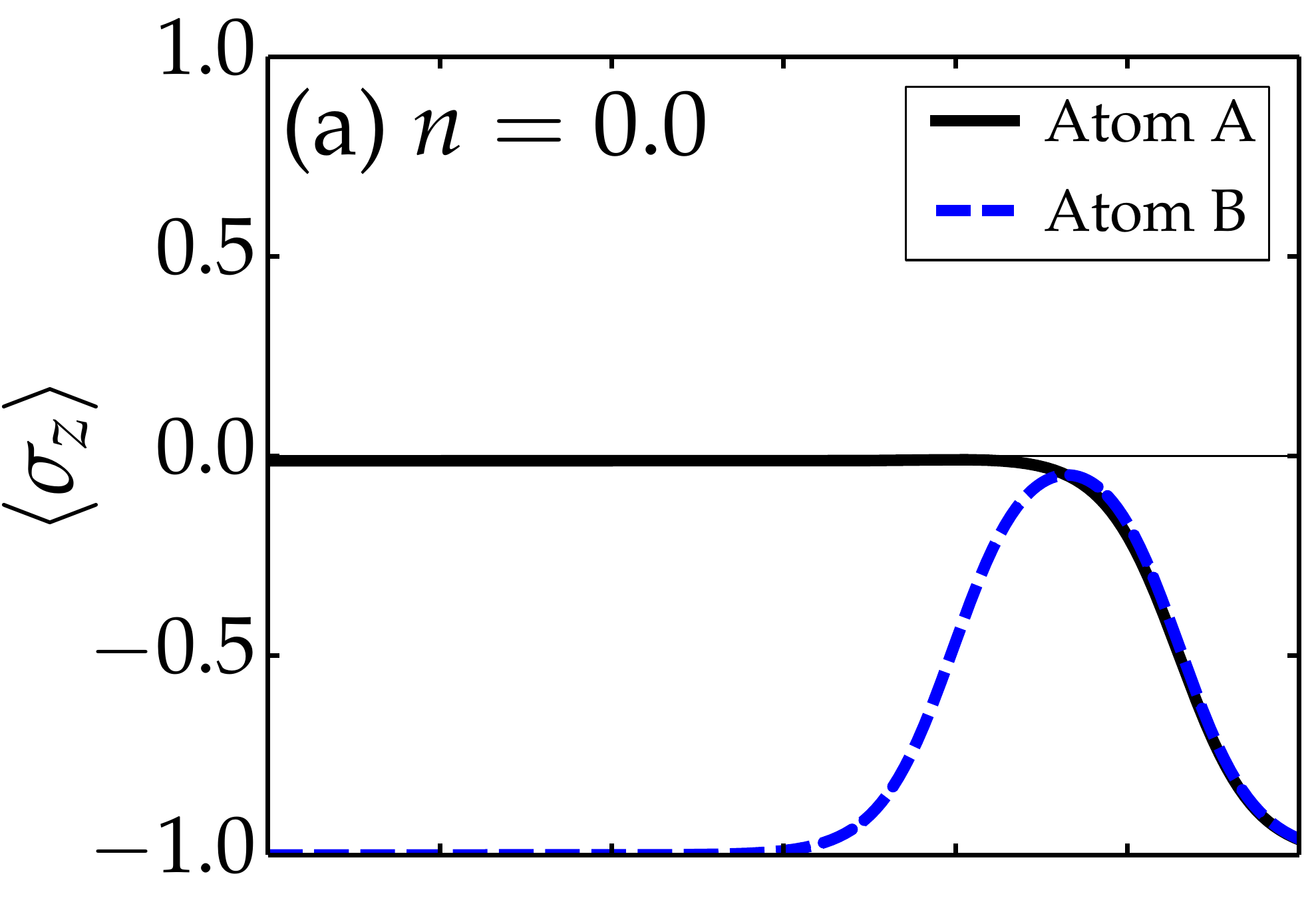} & \includegraphics[bb=50bp 0bp 559bp 395bp,scale=0.21]{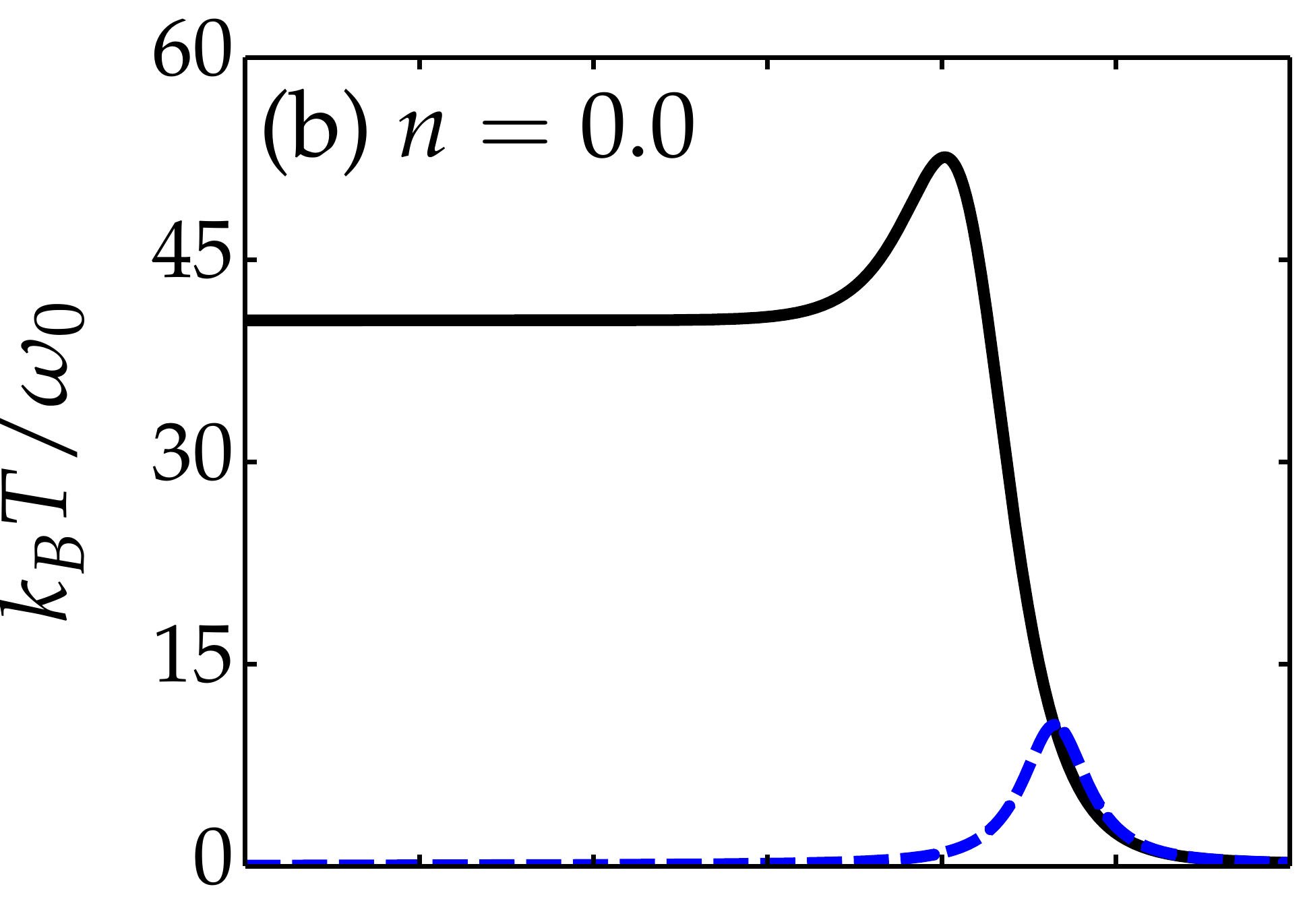}\tabularnewline
\includegraphics[bb=20bp 0bp 570bp 395bp,scale=0.21]{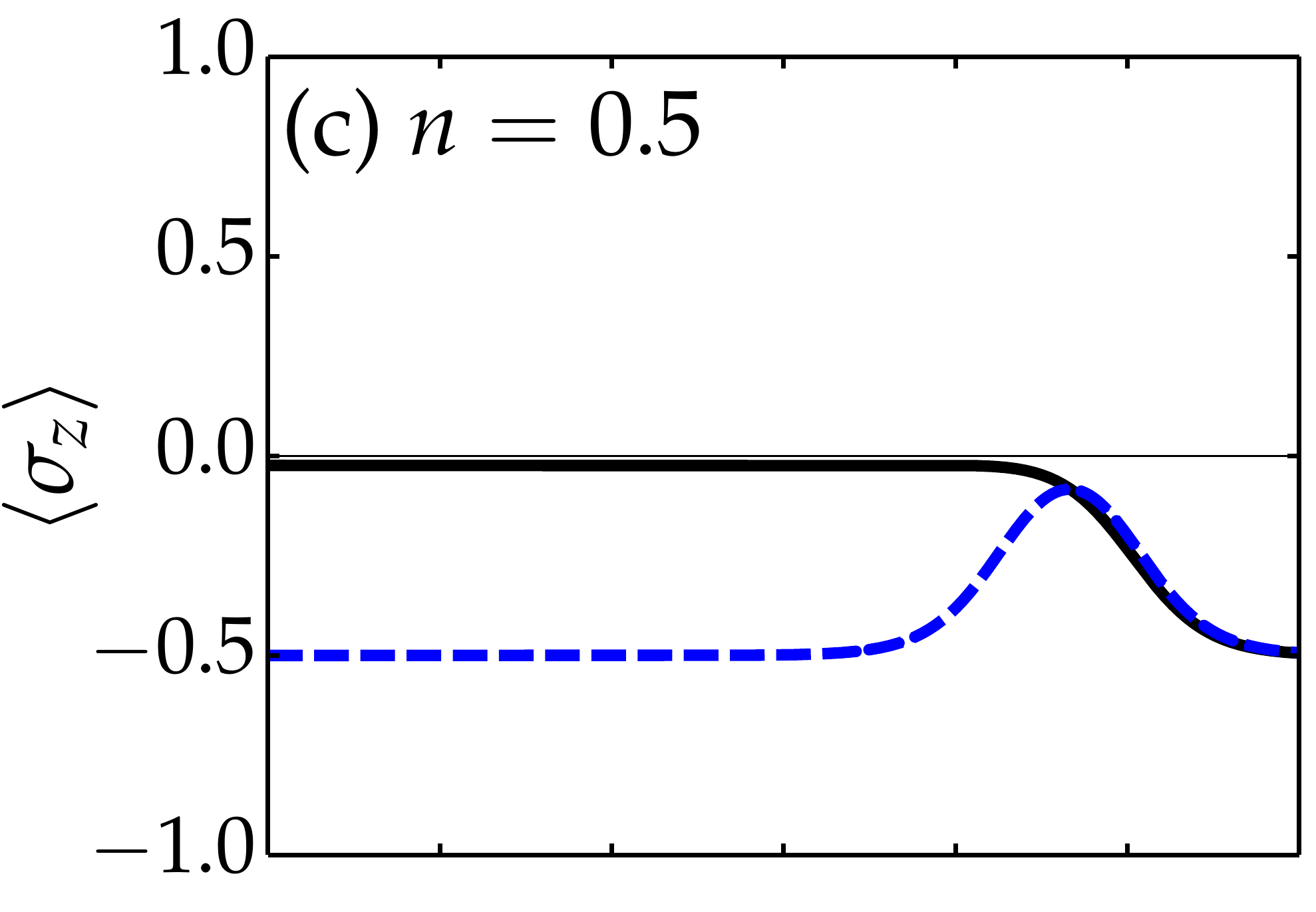} & \includegraphics[bb=50bp 0bp 559bp 395bp,scale=0.21]{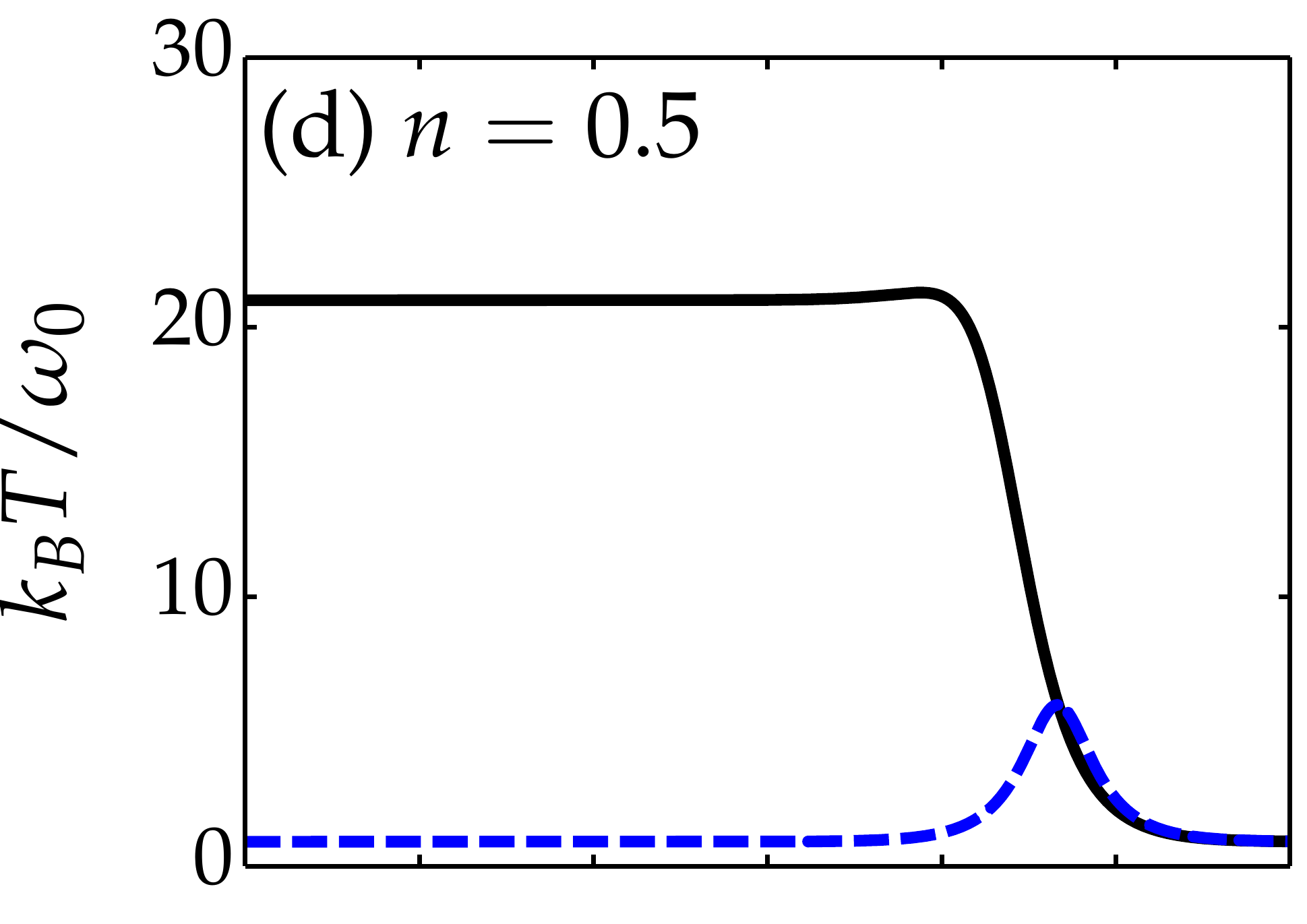}\tabularnewline
\includegraphics[bb=5bp 0bp 586bp 463bp,scale=0.21]{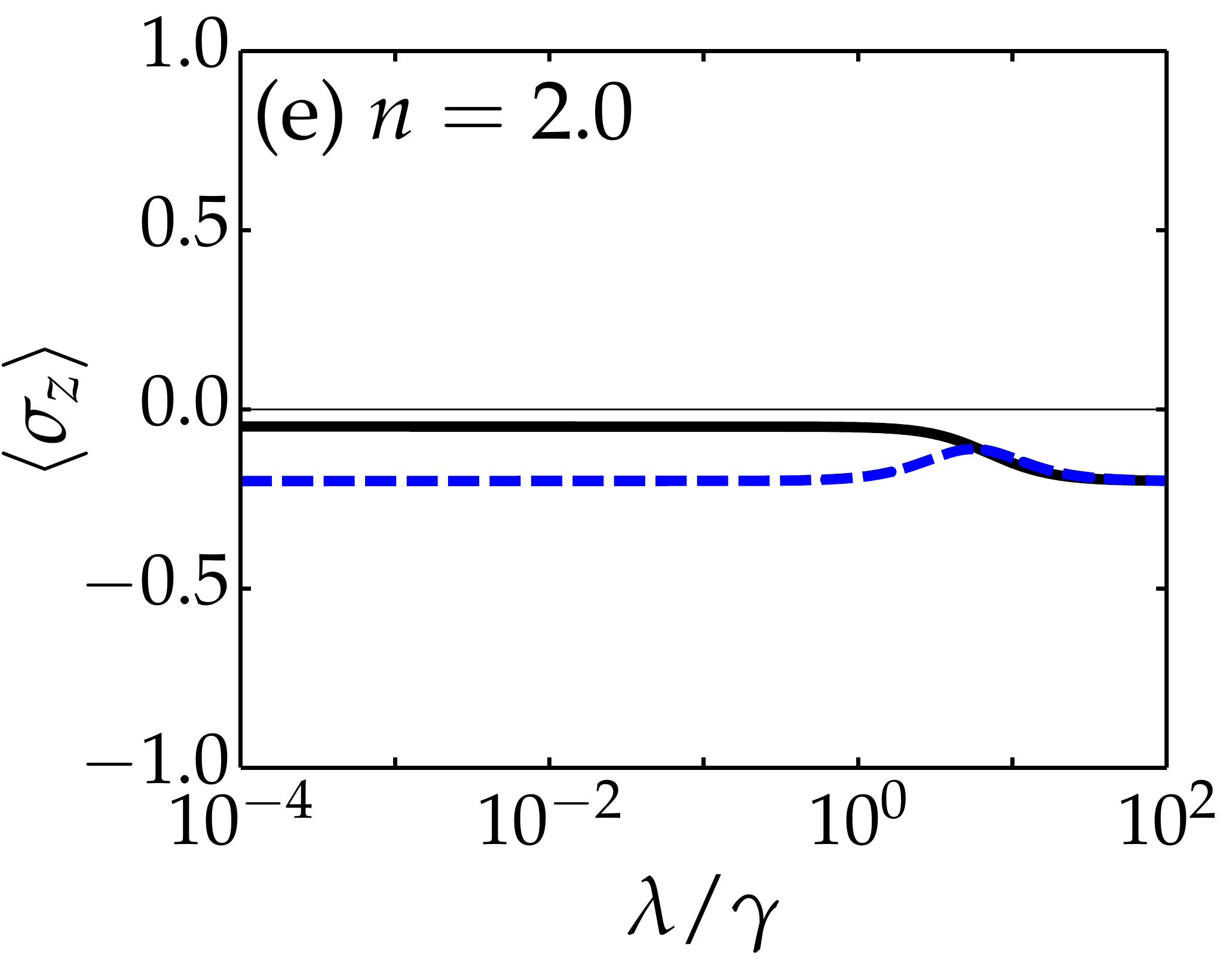} & \includegraphics[bb=20bp 0bp 586bp 463bp,scale=0.21]{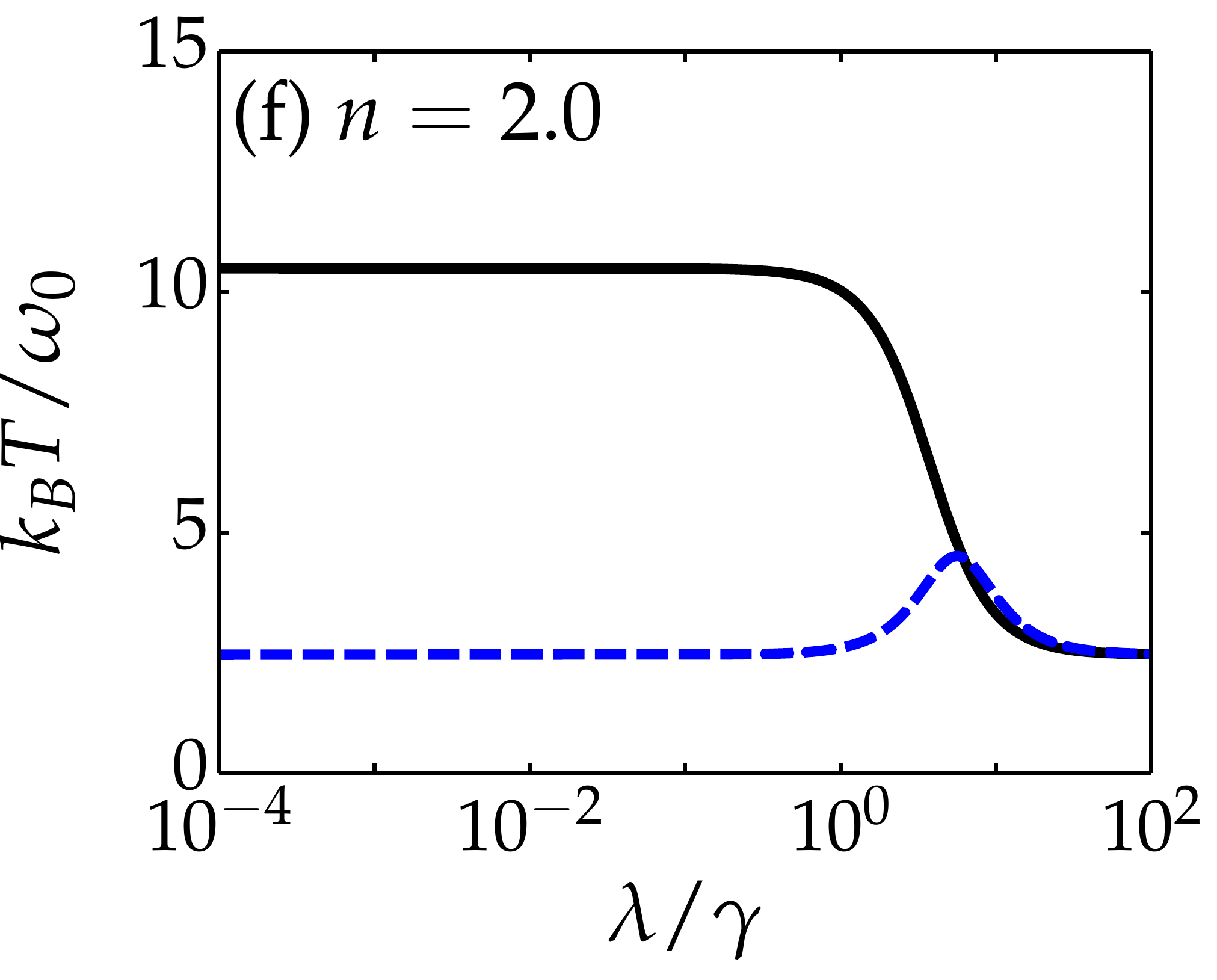}\tabularnewline
\end{tabular}\caption{\label{Fig4}Atom scaled average energy $\left\langle \sigma_{z}\right\rangle $
\emph{versus} scaled atom-atom coupling ($\lambda/\gamma$), Figs.
(a,c,e), and its scaled effective temperature $k_{B}T/\omega_{0}$
\emph{versus} $\lambda/\gamma$, Figs. (b,d,f), for $k=0$. The environment
thermal photon averages are $n=0.0$, $0,5$, and $2.0$, respectively.
Curves for atom A (B) is indicated by solid black (blue dashed) line.
Here we used $g_{k}=\sqrt{10}\gamma$, and $\gamma=\gamma_{A}=\gamma_{B}=\kappa$. }
\end{figure}

\begin{figure}[ptbh]
\centering{}%
\begin{tabular}{cc}
\includegraphics[bb=20bp 0bp 570bp 395bp,scale=0.21]{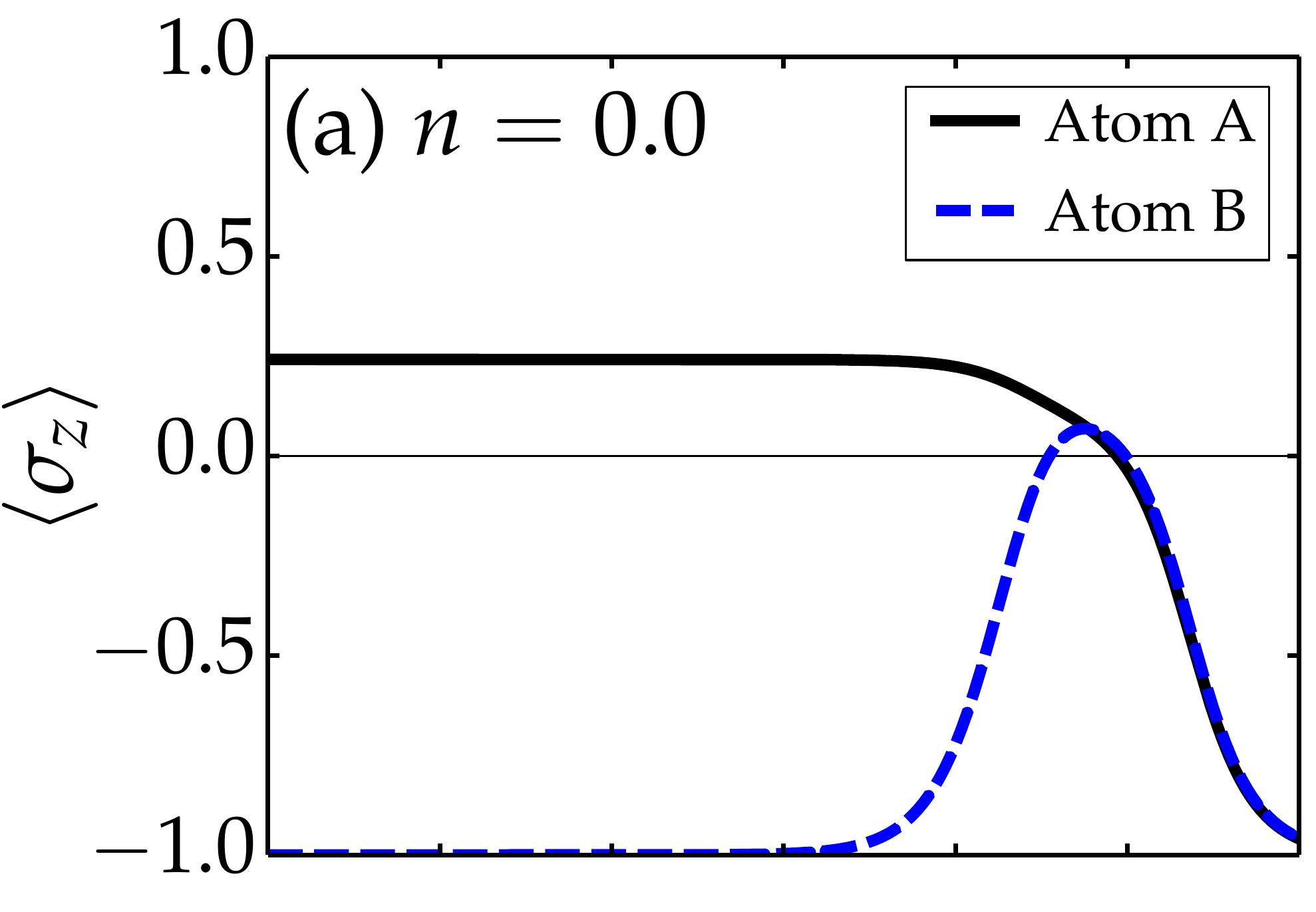} & \includegraphics[bb=50bp 0bp 560bp 395bp,scale=0.21]{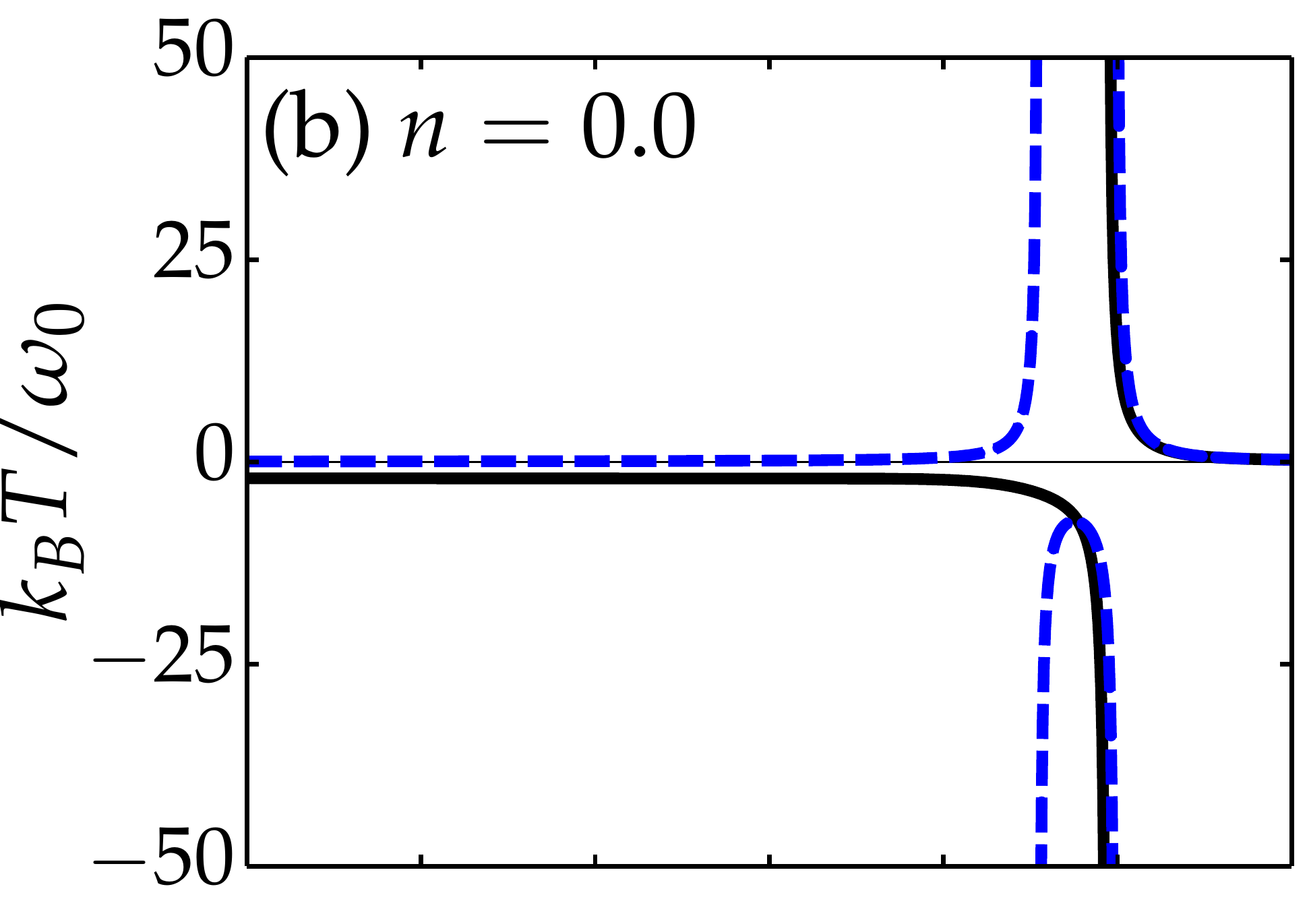}\tabularnewline
\includegraphics[bb=20bp 0bp 570bp 395bp,scale=0.21]{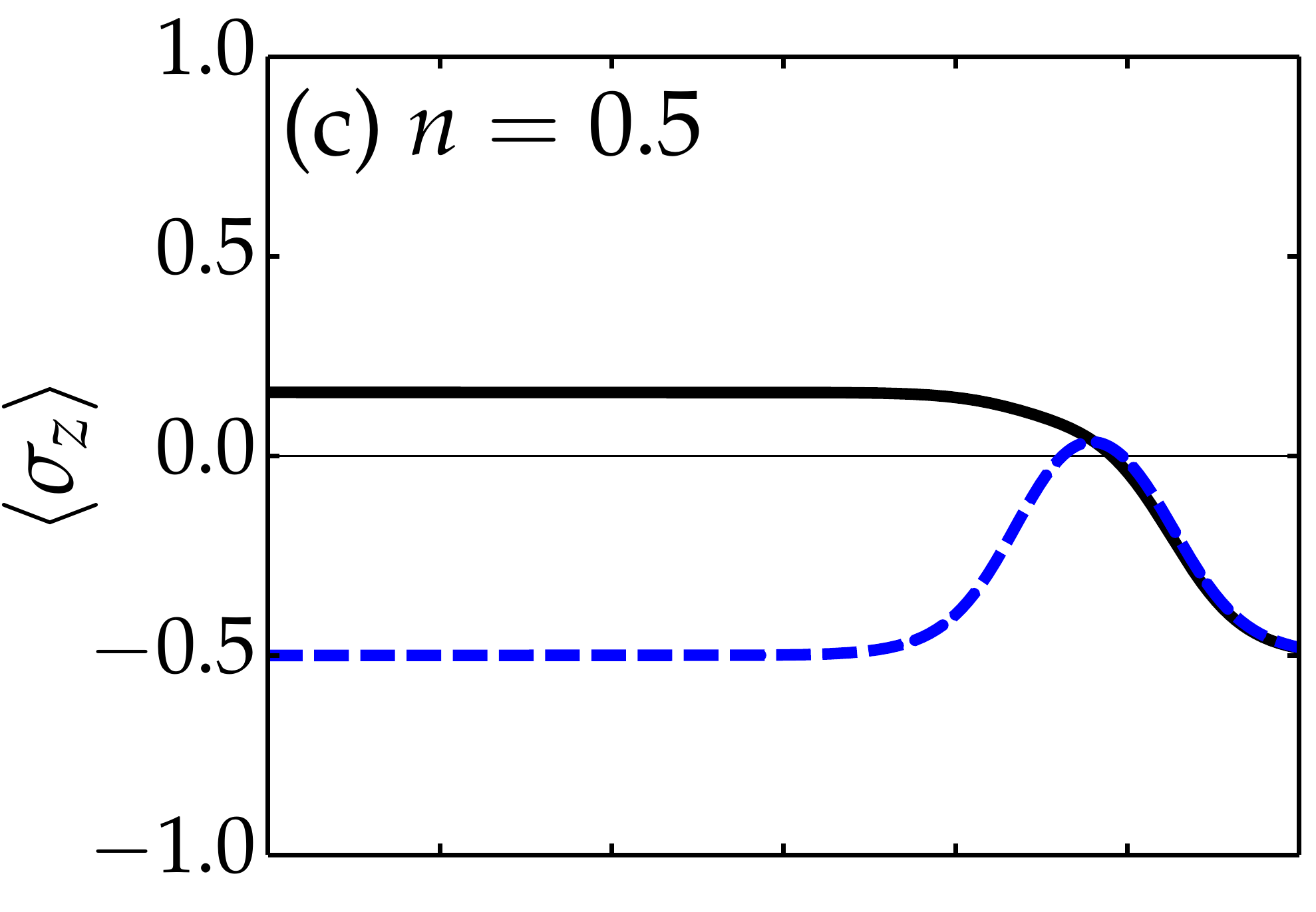} & \includegraphics[bb=50bp 0bp 560bp 395bp,scale=0.21]{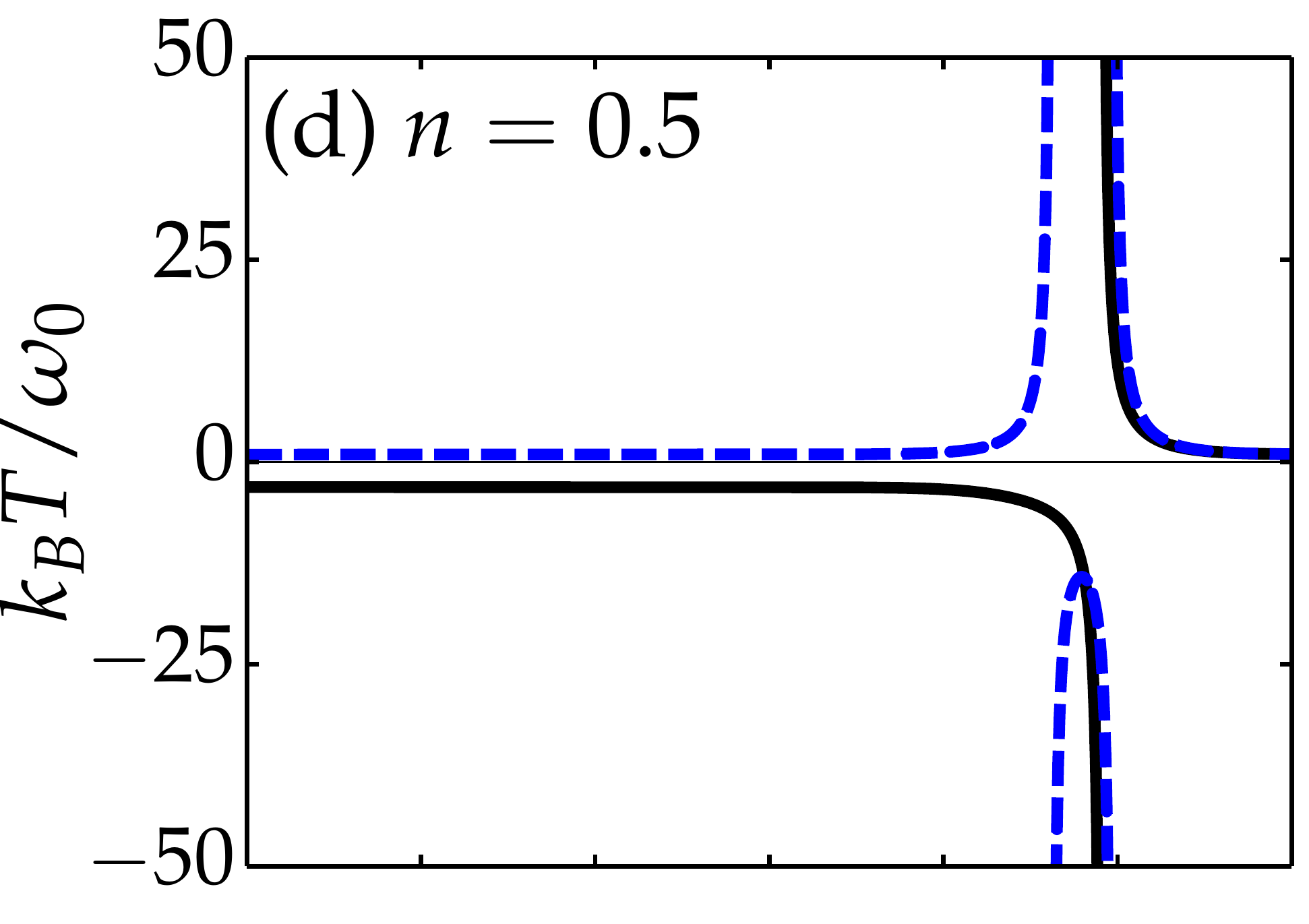}\tabularnewline
\includegraphics[bb=5bp 0bp 586bp 463bp,scale=0.21]{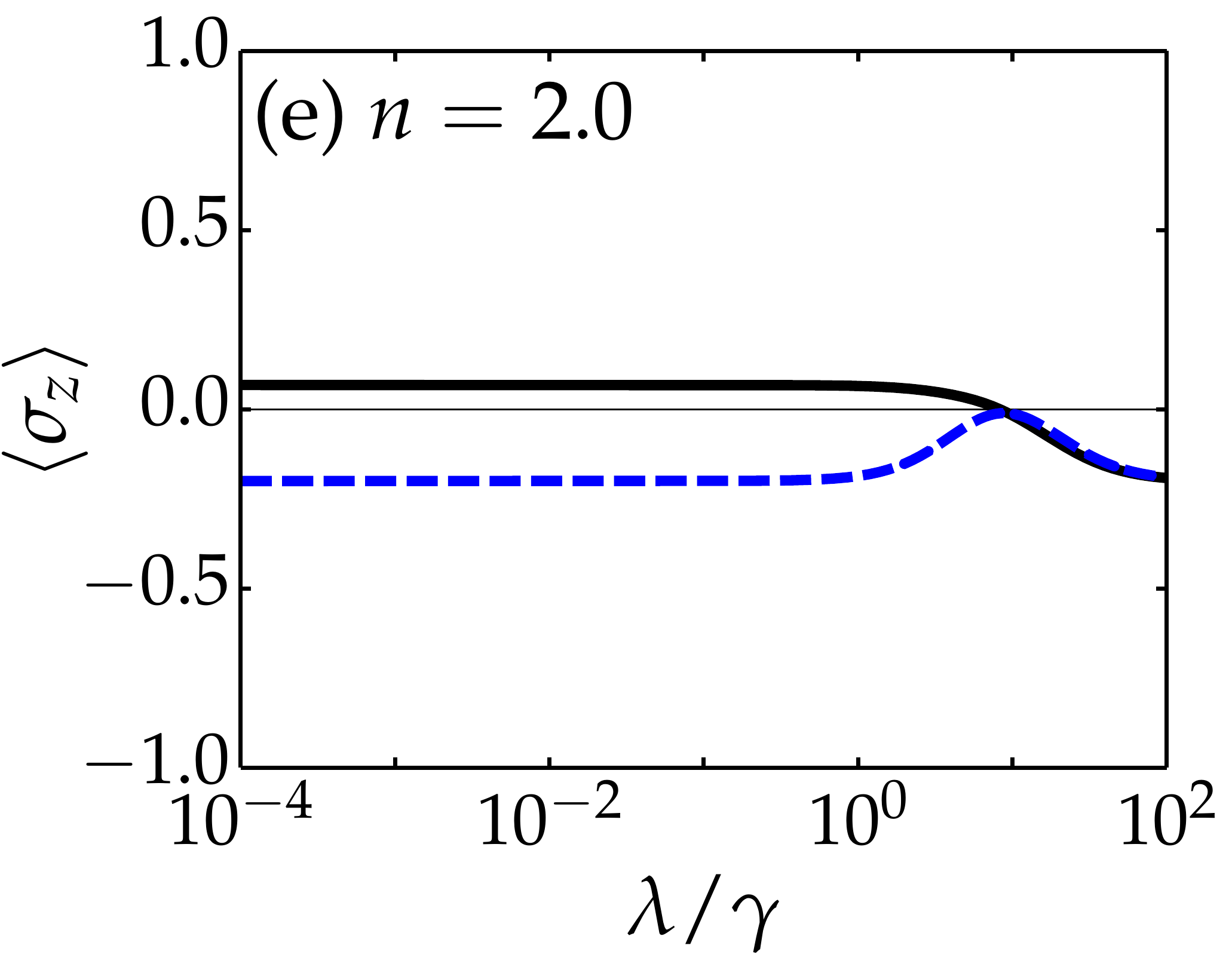} & \includegraphics[bb=20bp 0bp 586bp 463bp,scale=0.21]{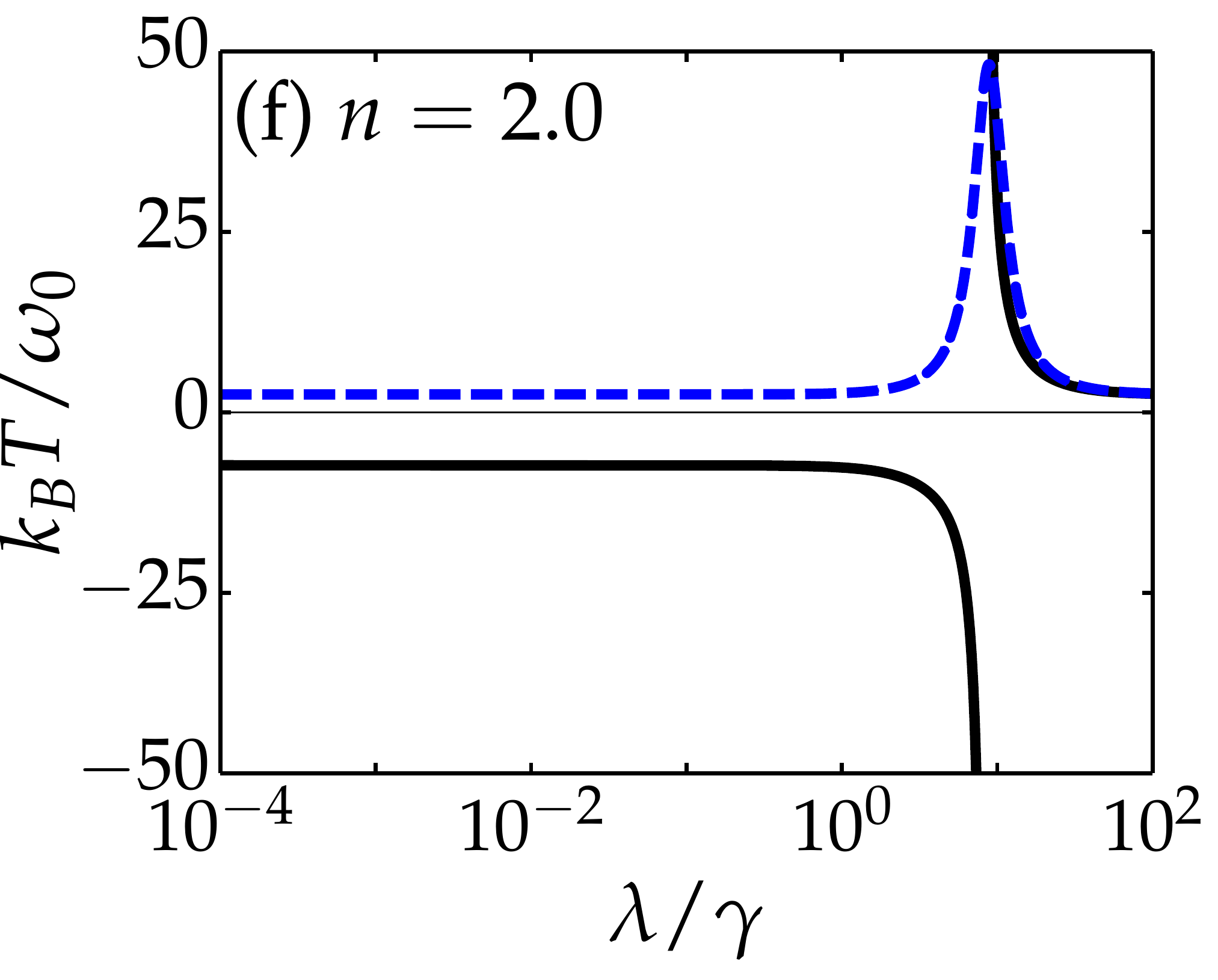}\tabularnewline
\end{tabular}\caption{\label{Fig5}Scaled average energy $\left\langle \sigma_{z}\right\rangle $\emph{
versus} scaled atom-atom coupling ($\lambda/\gamma$), Figs.(a,c,e),
and scaled effective temperature $k_{B}T/\omega_{0}$ \emph{versus}
$\lambda/\gamma$, Figs.(b,d,f), for $k=1$. The environment thermal
photon averages are $n_{th}=0.0$, $0.5$, and $2.0$, respectively.
Curves for atom A (B) is indicated by solid black (blue dashed) line.
Here we used $g_{k}=\sqrt{10}\gamma$ and $\gamma=\gamma_{A}=\gamma_{B}=\kappa$.}
\end{figure}

Now, let us recall that the cooperativity parameter comprehends the
coupling between ion $A$ and its vibrational mode as well as the
ion A decay rate and the vibrational mode damping through $C_{k}=g_{k}^{2}/\gamma_{A}\kappa$.
Thus, it is also interesting to study how the scaled average energy
and the effective temperature behaves\emph{ }when varying the strength
coupling $\lambda$ between atoms $\text{A}$ and B, and also asking
if it is possible to have thermalization between them. In Fig. \ref{Fig4}(a-f)
and \ref{Fig5}(a-f) we show the scaled average internal atom energy
$\left\langle \sigma_{z}\right\rangle $ and the effective temperature
\emph{versus} the scaled strength coupling $\lambda/\gamma$ for $k=0,1,2$,
and $3$ considering the environment thermal photon averages $n=0$,
$0.5$, and $2.0$. Note from Figs. \ref{Fig4}(a,c,e) that for $k=0$
the populations of atoms A and B are not inverted. Nevertheless, atom
B presents an interesting behavior: when the rate $\lambda/\gamma$
is increased, its population increases until reaching a maximum, and
then starts to decrease. On the other hand, atom A population always
decrease, until its population becomes equal to that of atom B, when
thermalization between atoms A and B thus occurs, see Figs. \ref{Fig4}(b,d,f),
for sufficiently high values of the rate $\lambda/\gamma$. Also it
is important to note that thermalization between atoms A and B may
occur at different temperatures to that of their environments, as
can be seen from Figs. \ref{Fig4}(b,d,f), where the environment has
temperatures corresponding to $n=0,0.5$ and $2.0$.

In Figs. \ref{Fig5}(a-f) we now study the model $k=1$. Now, different
from Fig.\ref{Fig4}, both atoms can present negative temperatures.
Note that atom B energy increases until crossing the zero energy line,
thus inverting its population and acquiring negative temperature,
and then starts to decrease until its energy crosses back the zero
energy line, acquiring positive temperature. On the contrary, atom
A, which starts with inverted population, diminishes its internal
energy until crossing the zero line energy, acquiring a positive temperature
for sufficiently high value of the rate $\lambda/\gamma$. Also, as
in the previous case, for sufficiently high values of the rate $\lambda/\gamma$
both atoms A and B thermalize with each other, although not thermalizing
with their environment. It is interesting to note that atom A and
B \emph{can thermalize even at negative temperatures}, as is better
seen in Figs. \ref{Fig5}(a-d). The role of the atoms A and B environments
is to diminish population inversion, see the blue-dash line in Figs.\ref{Fig5}(a,c,e),
and thus the negative temperature effect. For the parameters used
here, the negative temperature of atom B, blue-dash line, is completely
suppressed at a temperature corresponding to an average thermal photon
$n=2.0$. 

\begin{figure}[ptbh]
\centering{}%
\begin{tabular}{cc}
\includegraphics[bb=20bp 0bp 570bp 395bp,scale=0.21]{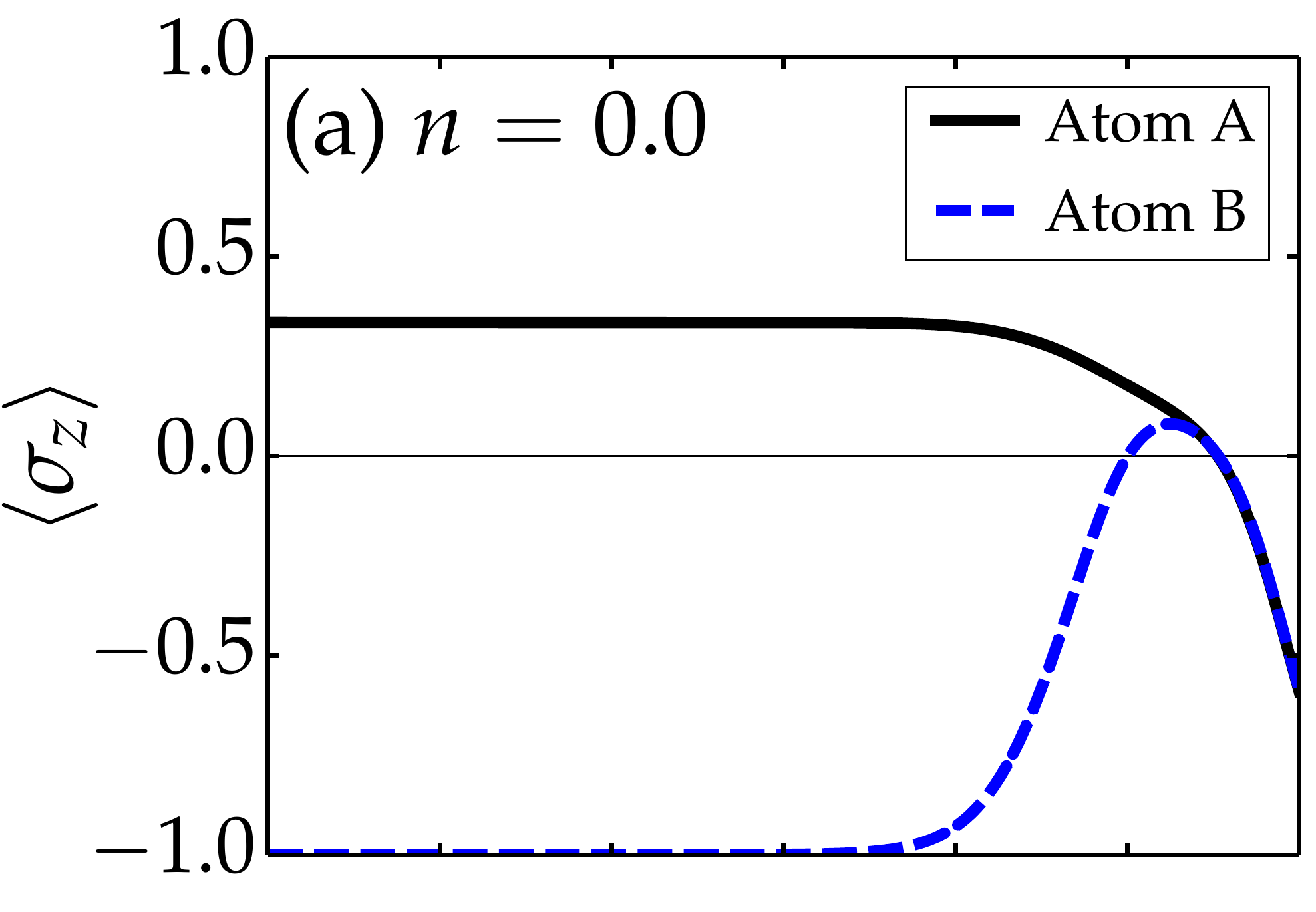} & \includegraphics[bb=50bp 0bp 560bp 395bp,scale=0.21]{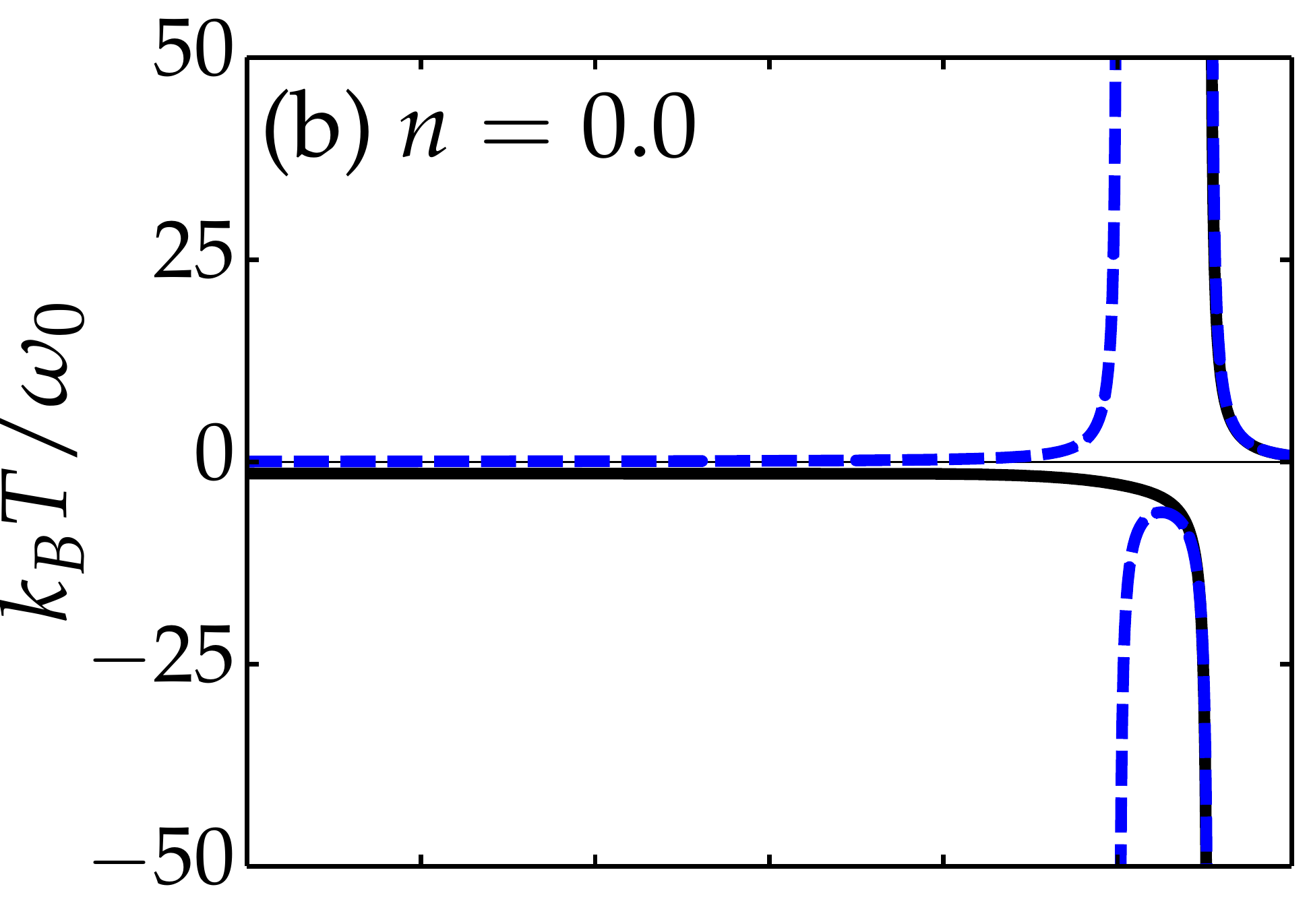}\tabularnewline
\includegraphics[bb=20bp 0bp 570bp 395bp,scale=0.21]{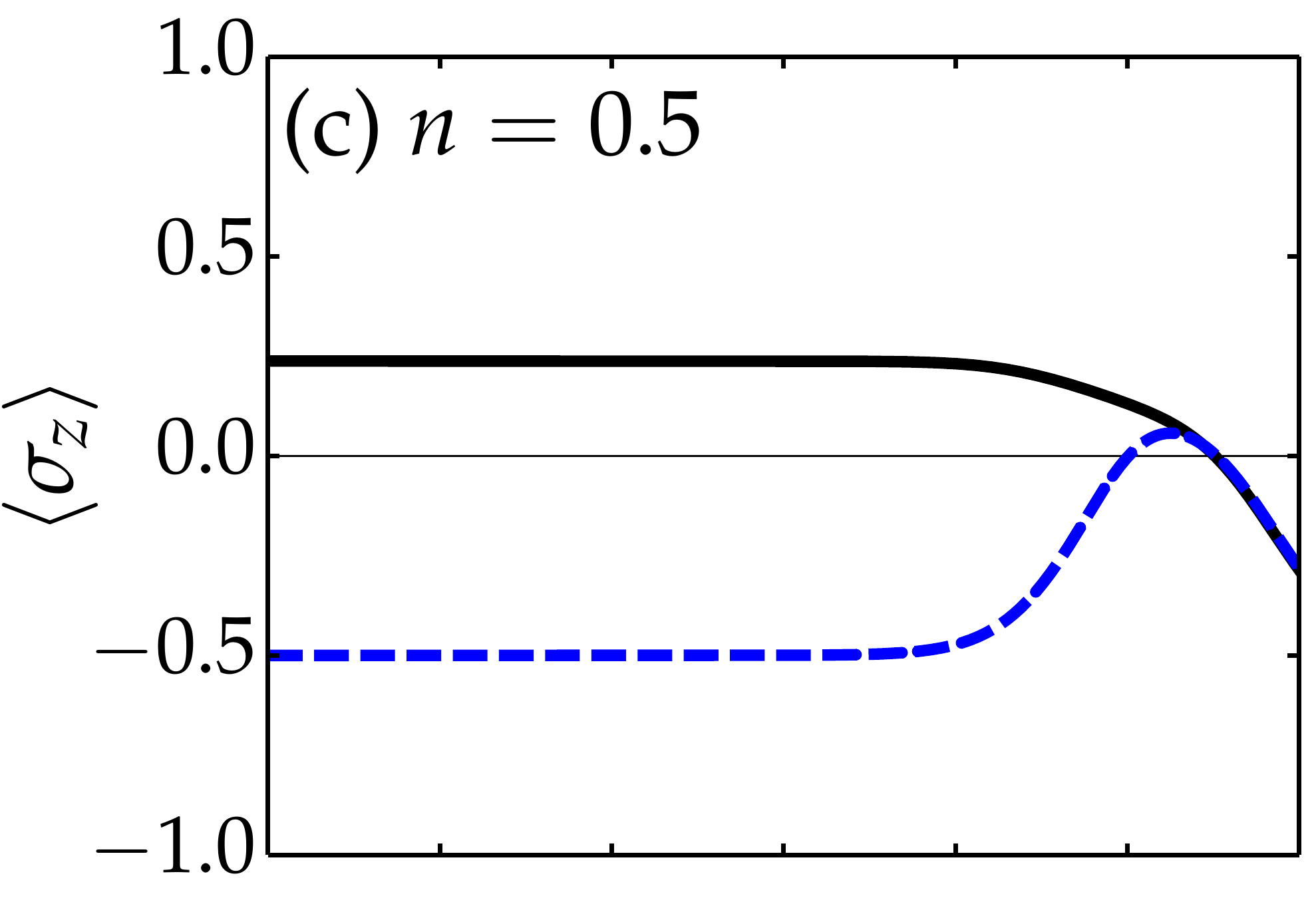} & \includegraphics[bb=50bp 0bp 560bp 395bp,scale=0.21]{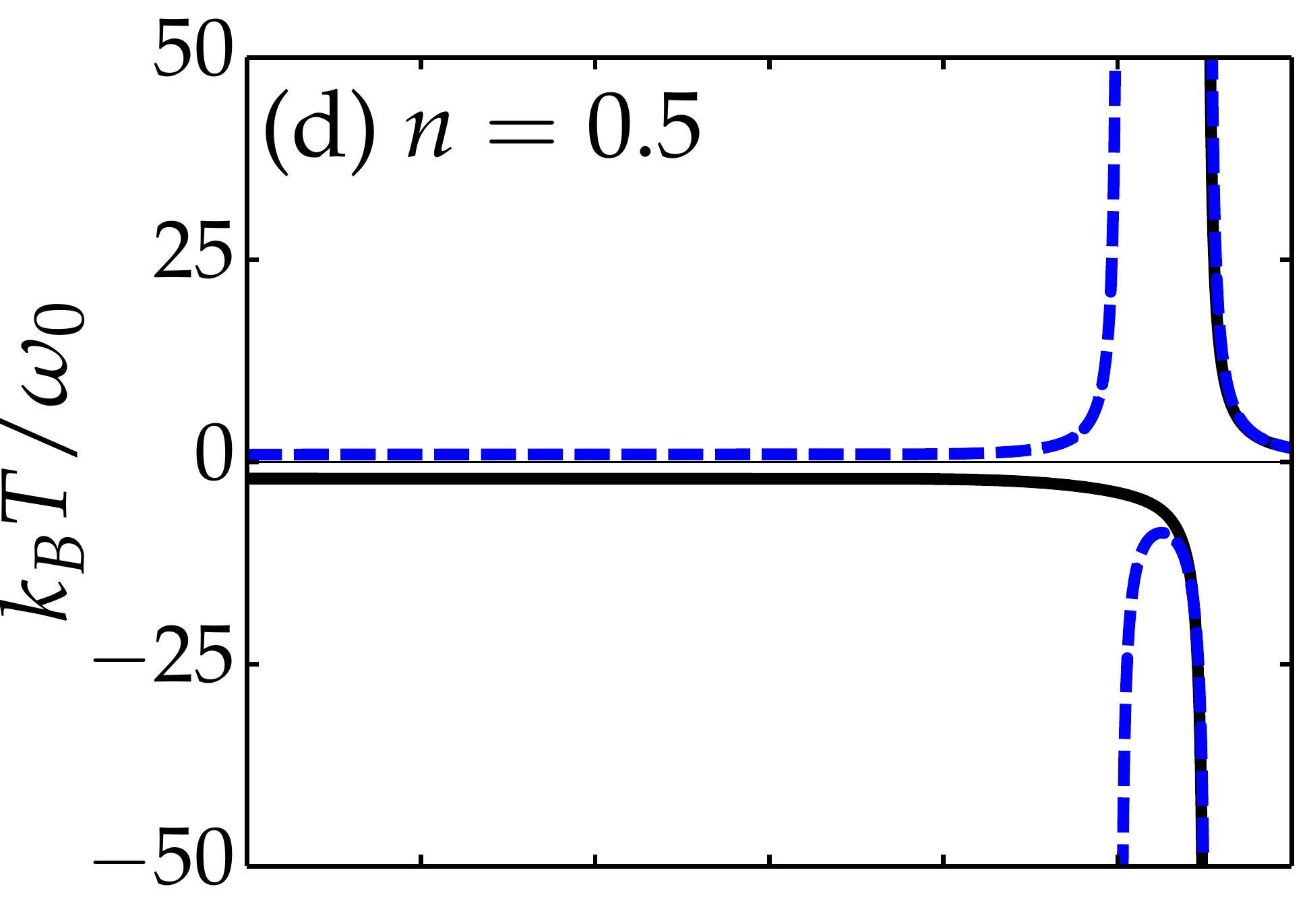}\tabularnewline
\includegraphics[bb=5bp 0bp 586bp 463bp,scale=0.21]{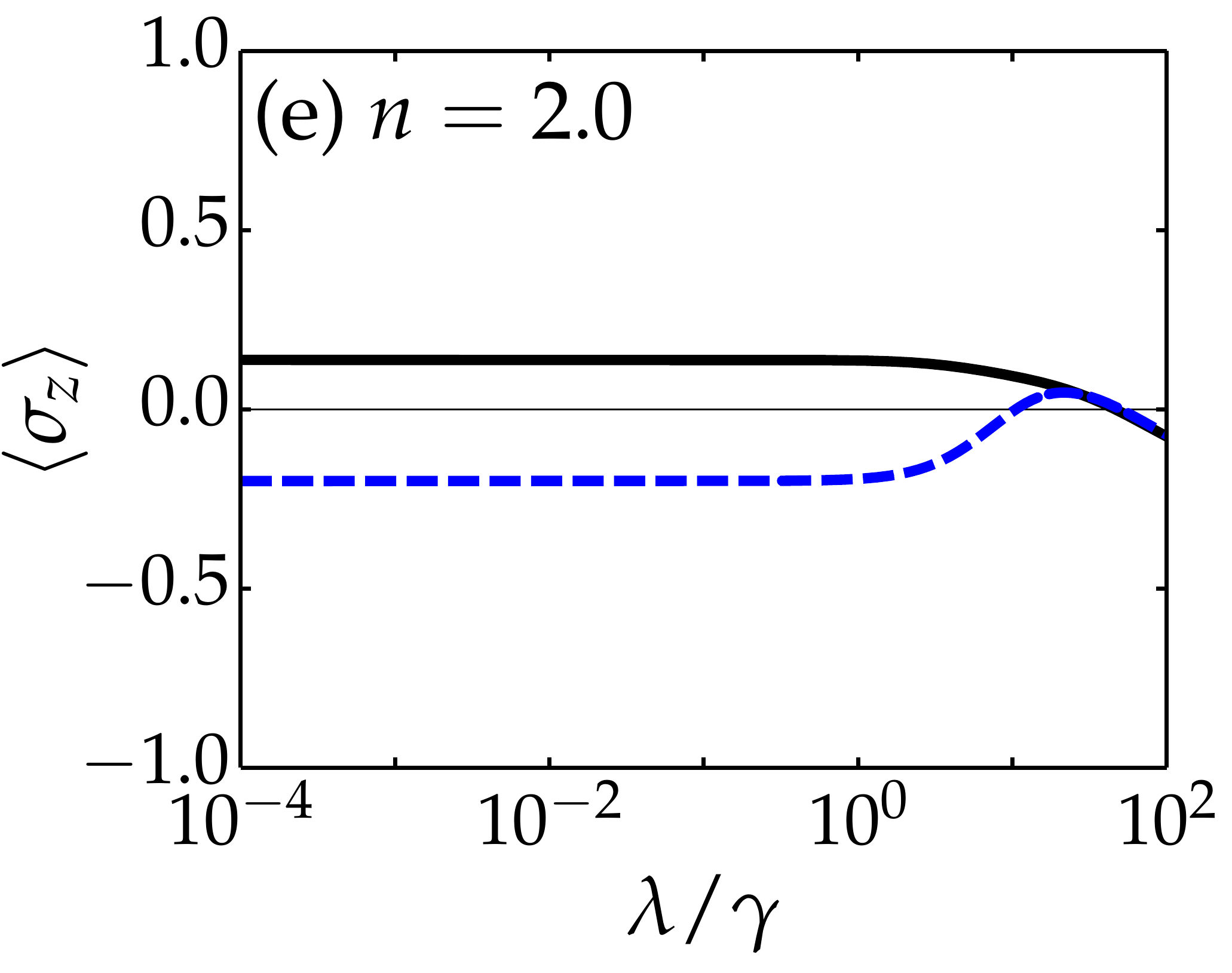} & \includegraphics[bb=20bp 0bp 586bp 463bp,scale=0.21]{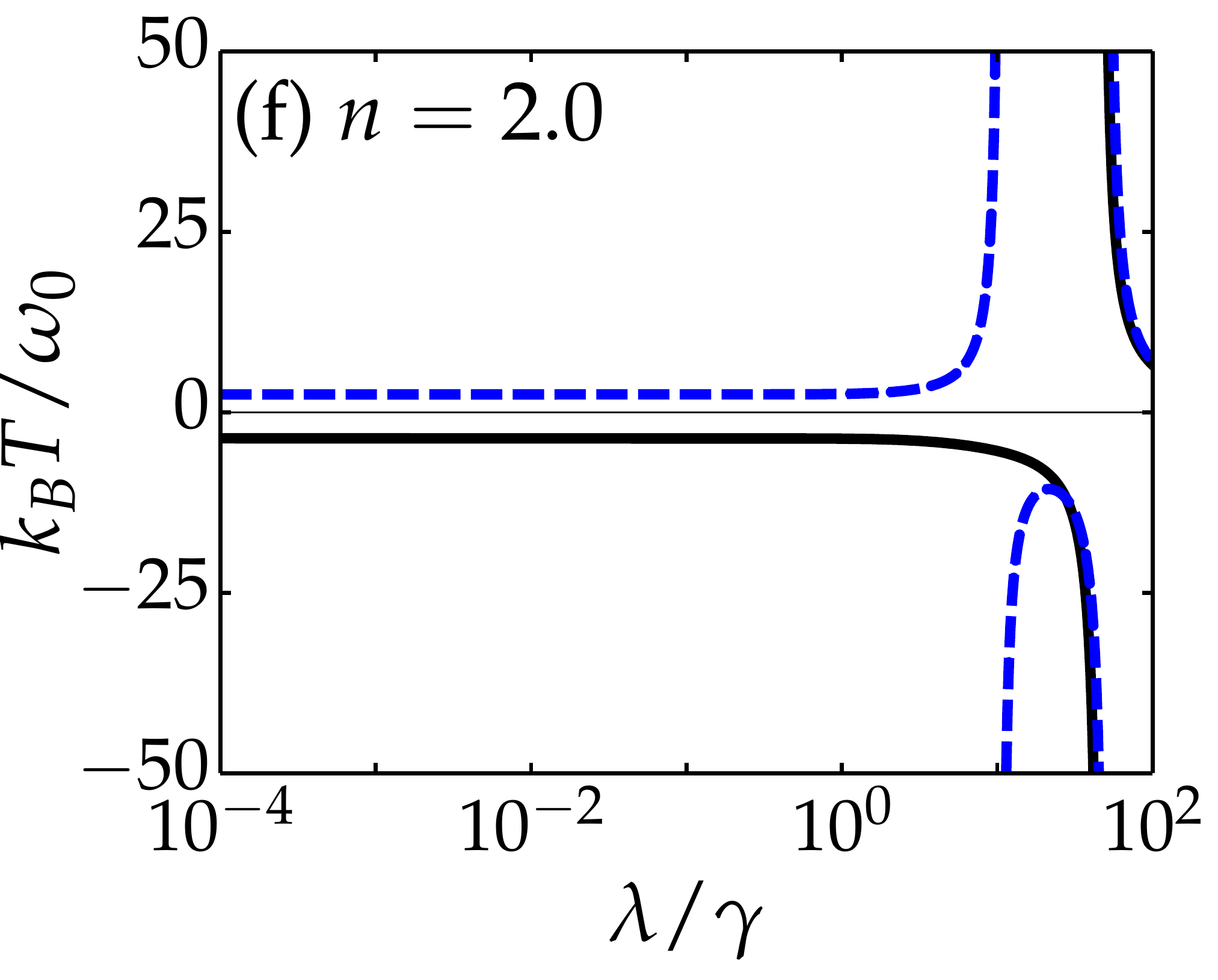}\tabularnewline
\end{tabular}\caption{\label{Fig6}Scaled average energy $\left\langle \sigma_{z}\right\rangle $\emph{
versus} cooperativity, Figs. (a,c,e), and scaled effective temperature
$k_{B}T/\omega_{0}$ \emph{versus} cooperativity, Figs. (b,d,f), for
$k=2$. The environment thermal photon averages are $n=0.0$, $0.5$,
and $2$, respectively. Curves for atom A (B) is indicated by solid
black (blue dashed) line. Here we used $g_{k}=\sqrt{10}\gamma$ and
$\gamma=\gamma_{A}=\gamma_{B}=\kappa$.}
\end{figure}

Figs. \ref{Fig6} (a-f) now take into account the Hamiltonian model
$k=2$. The same pattern is observed, as in the previous Figs. (\ref{Fig4}-\ref{Fig5}):
the stronger the nonlinearity of the Hamiltonian model, the higher
the population inversion obtained for both atoms. Now the maximum
environment temperature for both atoms is not enough to completely
suppress negative temperature for atom B, as can clearly be seen from
Fig.\ref{Fig6}(e,f). Also, thermalization between atoms A and B may
be seen even for negative temperatures for certain values of the rate
$\lambda/\gamma$ and, by increasing this rate a bit, both atoms acquire
positive temperature, where thermalization also occurs. Our numerical
analyses show that the same qualitative behavior is seen for the model
$k=3$ (not shown).

\section{Unconventional cooling by heating (CBH)}

\begin{figure}[ptbh]
\centering{}%
\begin{tabular}{cc}
\includegraphics[bb=20bp 0bp 570bp 395bp,scale=0.21]{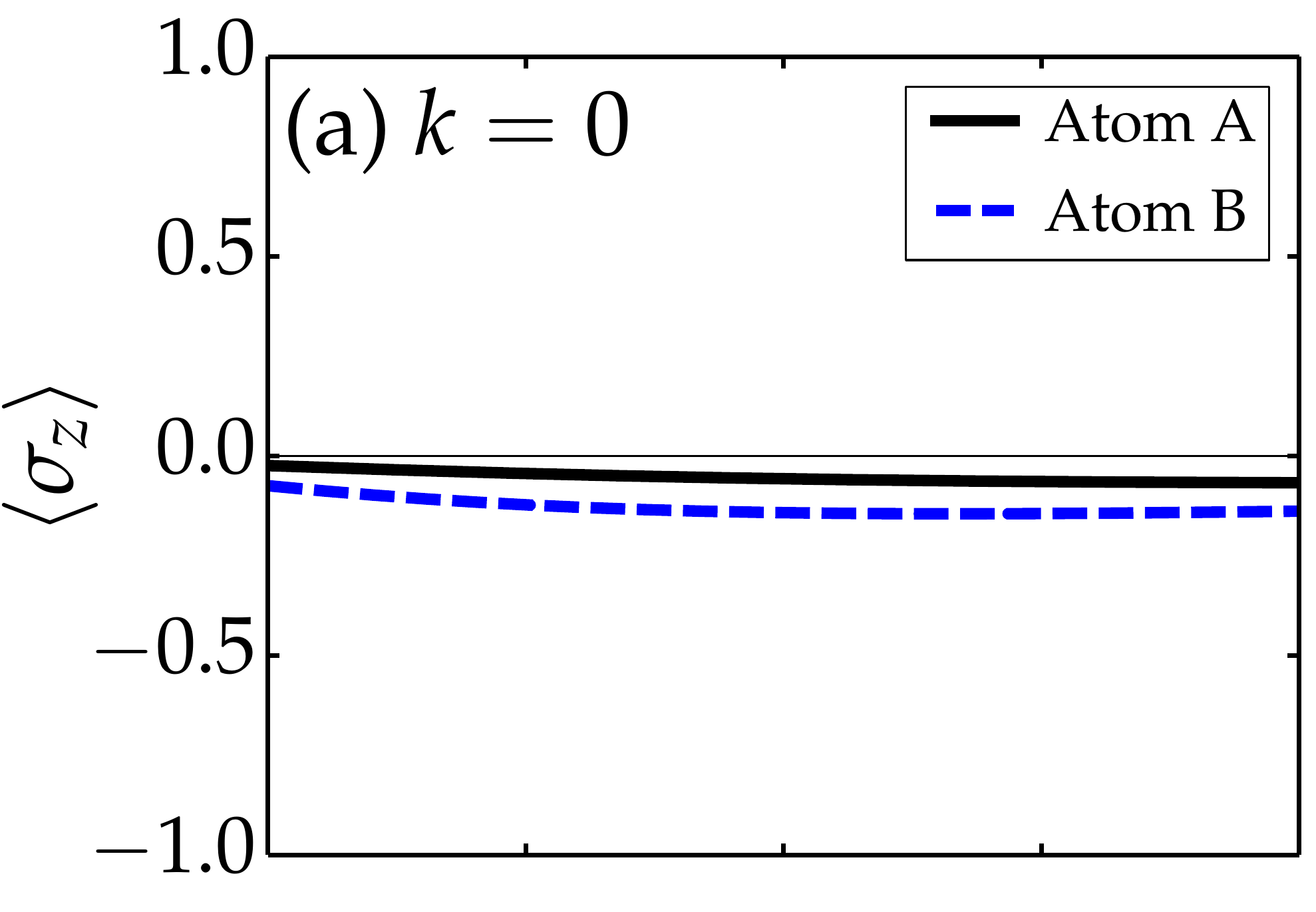} & \includegraphics[bb=50bp 0bp 559bp 395bp,scale=0.21]{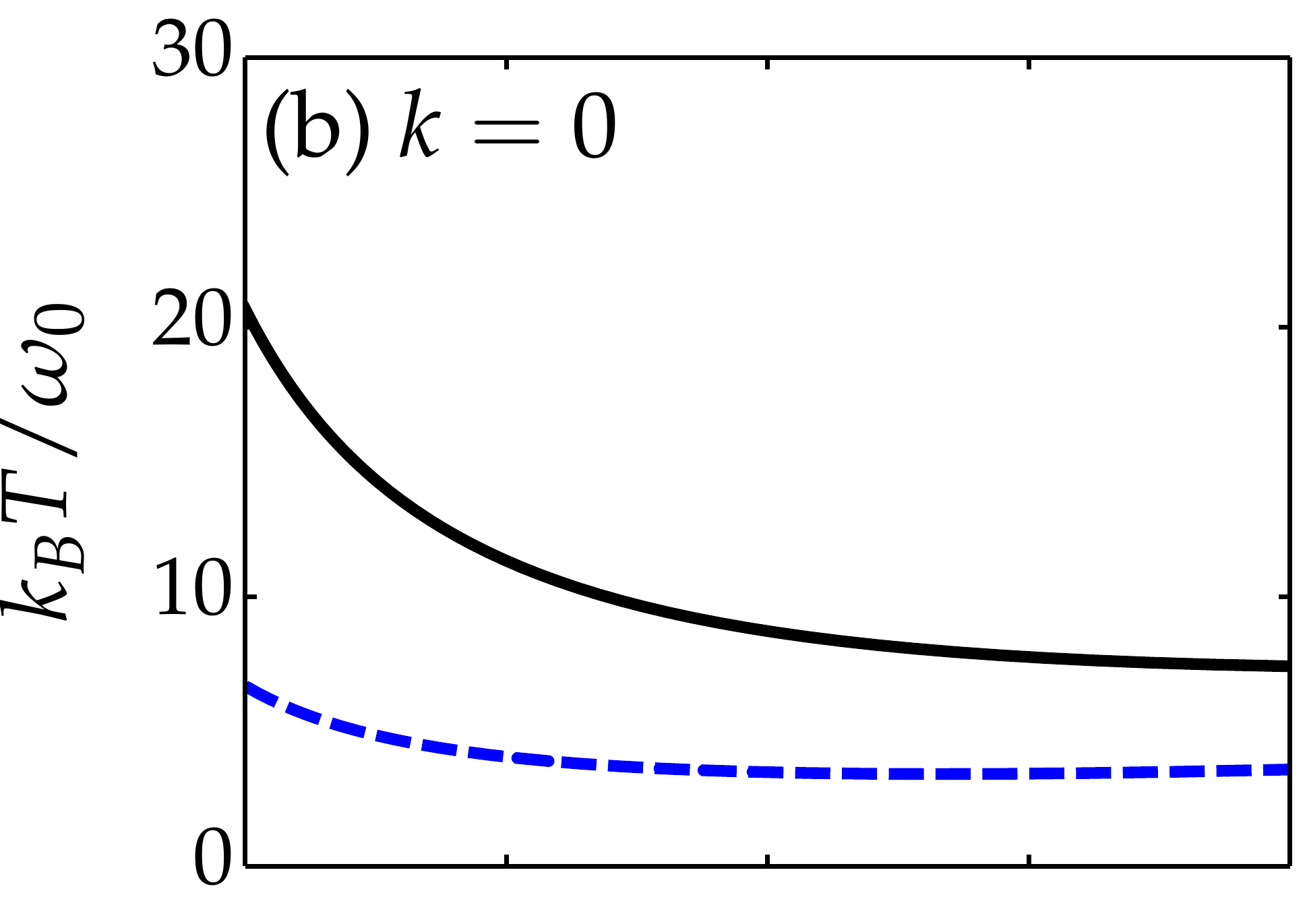}\tabularnewline
\includegraphics[bb=20bp 0bp 570bp 395bp,scale=0.21]{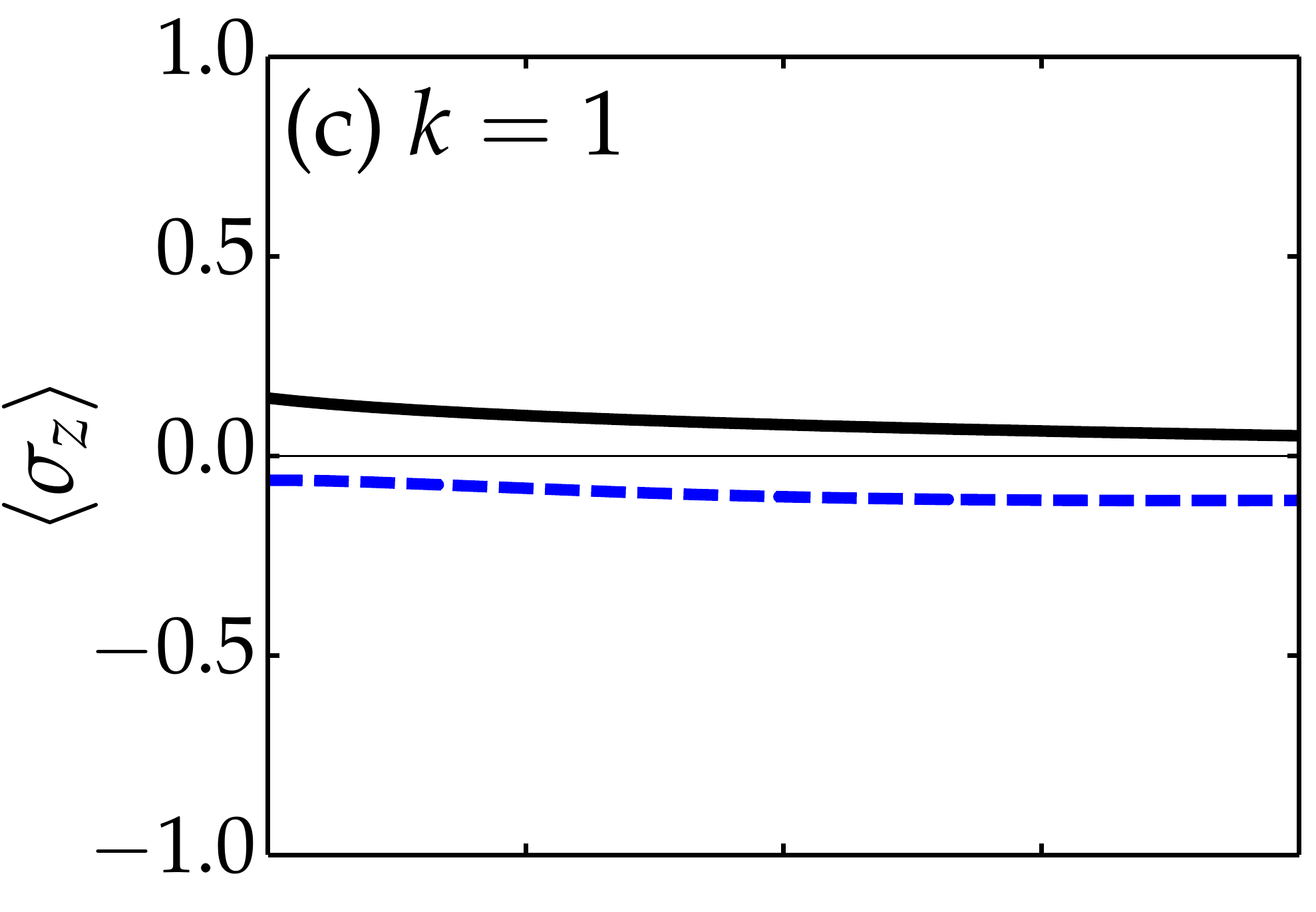} & \includegraphics[bb=50bp 0bp 560bp 395bp,scale=0.21]{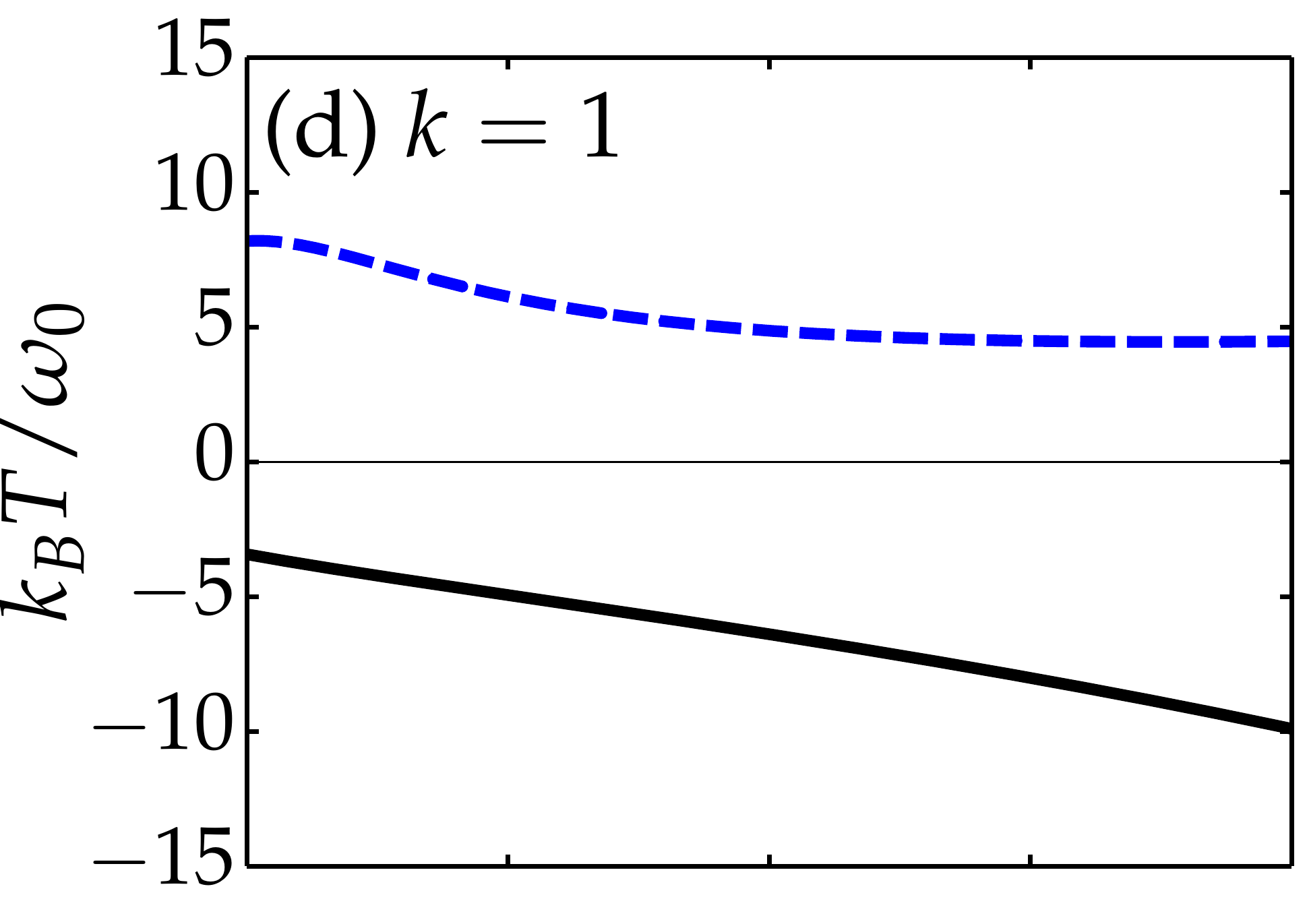}\tabularnewline
\includegraphics[bb=20bp 0bp 570bp 395bp,scale=0.21]{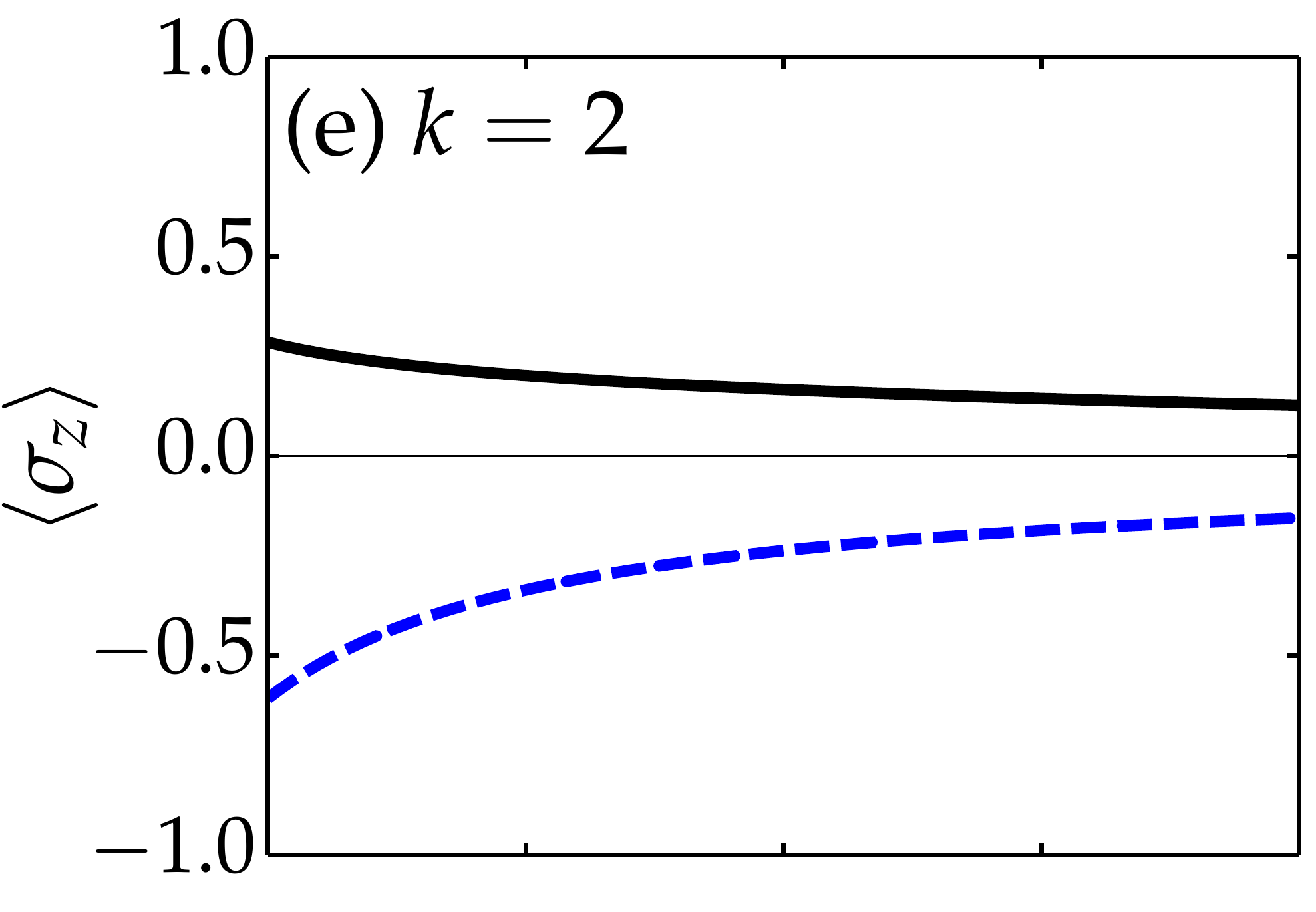} & \includegraphics[bb=50bp 0bp 560bp 395bp,scale=0.21]{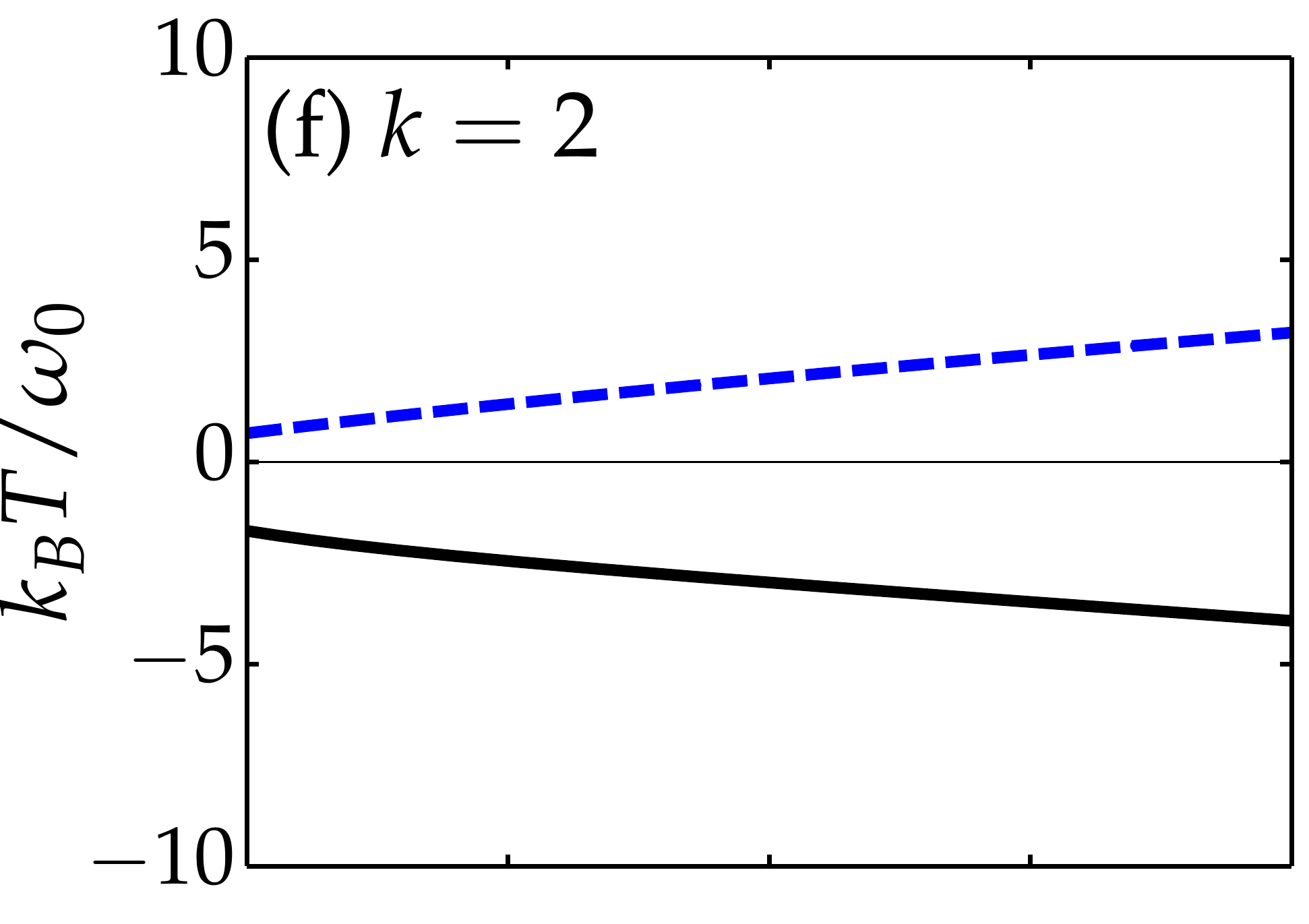}\tabularnewline
\includegraphics[bb=5bp 0bp 586bp 463bp,scale=0.21]{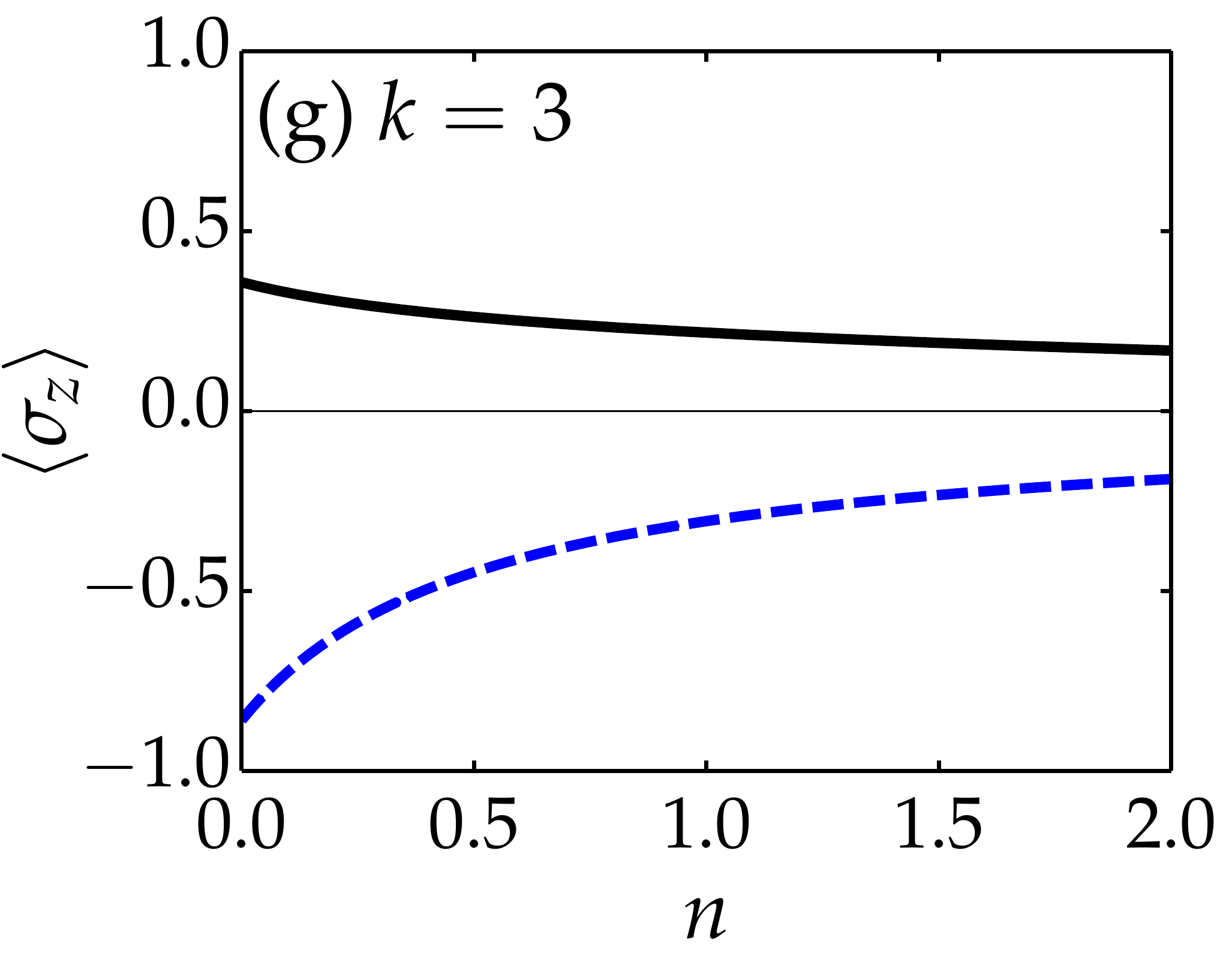} & \includegraphics[bb=20bp 0bp 586bp 463bp,scale=0.21]{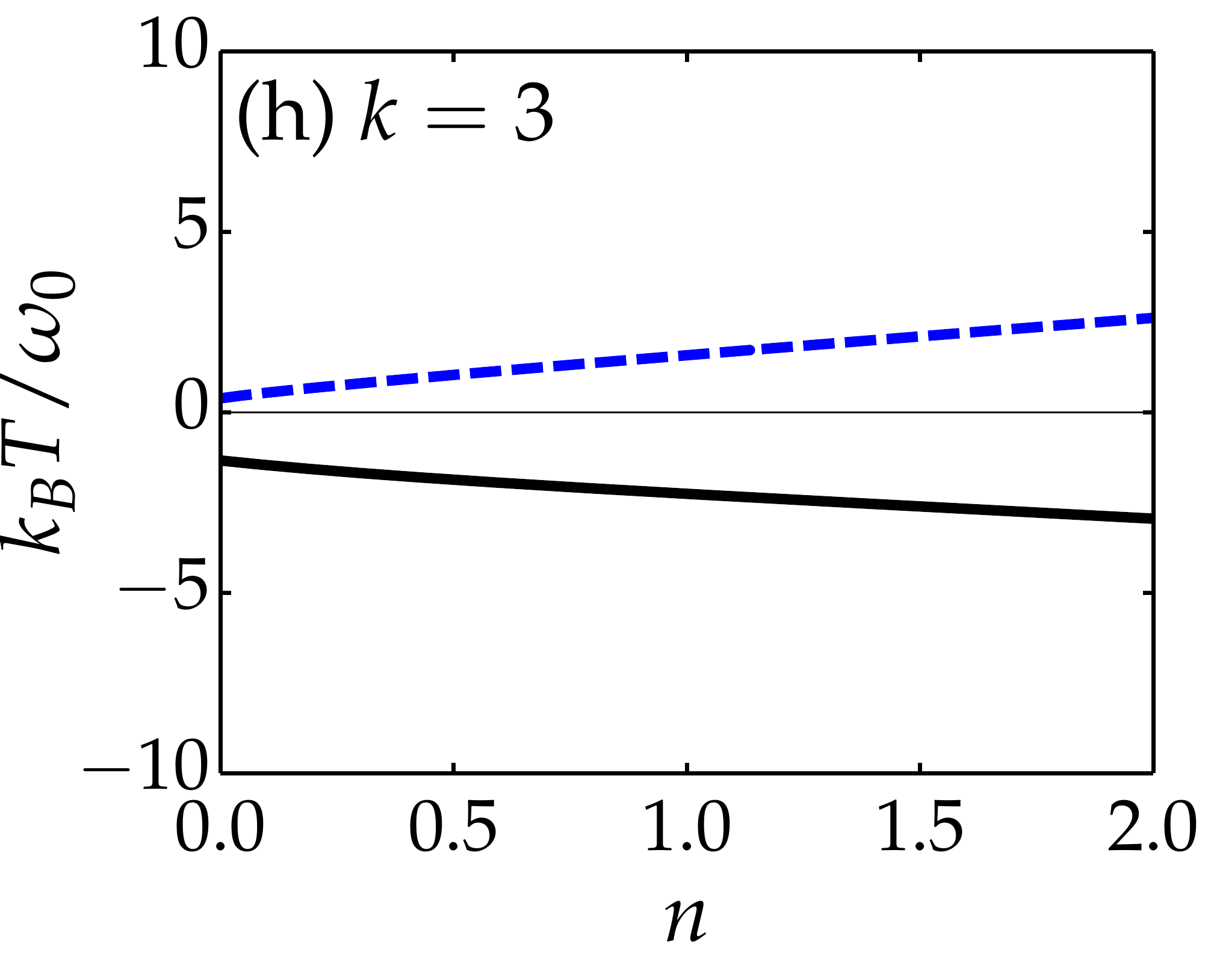}\tabularnewline
\end{tabular}\caption{\label{Fig7}The atom scaled average energy $\left\langle \sigma_{z}\right\rangle $
and its scaled effective temperature $k_{B}T/\omega_{0}$ \emph{versus}
the environment thermal photon average $n$ for the models $k=0,1,2,3$
and atoms A (solid black line) and B (blue dashed line). Except by
$k=0$, all the other models present negative temperature for atom
A (solid black line). Here we used $g_{k}=\sqrt{10}\gamma$ and $\lambda=3\gamma$. }
\end{figure}

Now let us analyze the case where both the bosonic mode and atoms
A and B are surrounded by an environment at the same temperature $T$
characterized by an average number $n$ of thermal photons. We again
assume here that both atoms decay with the same rate $\gamma_{A}=\gamma_{A}=\kappa=\gamma$
and interact with each other through the natural coupling $H_{N}=\lambda(\sigma_{+}^{A}\sigma_{-}^{B}+\sigma_{-}^{A}\sigma_{+}^{B})$.
The cooperativity is fixed at $C=10$\textbf{.} Meanwhile, atom A
is pumped with a laser leading to the effective Hamiltonian Eq. \eqref{AJC}
between atom A and the quantized vibrational mode. Since in this case
the interaction considered between the two atoms is the usual (or
natural) one, one could expect thermalization between the atoms, which
would eventually prevent CBH \cite{Rossato12}, at least in conventional
systems with positive temperatures. Actually, as discussed, e.g. in
Refs. \cite{Rossato12}, to achieve CBH in conventional systems (positive
temperatures) it is necessary to engineer a Hamiltonian whose major
contribution is due to counter-rotating terms. This is because two
systems, modeled by the usual Hamiltonians as given by matter-radiation
interaction, generally thermalize with their environment and, when
in contact, thermalize with each other.\textbf{ }However, as we saw
in Figs. \eqref{Fig2} and \eqref{Fig3}, the steady state of atom
A can display negative temperature, and thus an \emph{unconventional
CBH} with systems presenting negative temperatures can indeed occur,
despite the coupling being a natural (not engineered) one. To see
this, in Fig. \eqref{Fig7}(a-h) we show the scaled average internal
atom energy and the scaled effective temperature \emph{versus} the
environment thermal photon average $n$ for the models $k=0,1,2,3$.
It is to be noted that when $k=0$ Figs. \eqref{Fig7}(a,b), the populations
of the atoms A and B are not inverted, and thus we have the \emph{conventional}
(positive temperatures) CBH: by increasing its environment temperature,
its internal energies, and therefore its effective temperature, is
diminished. On the other hand, unconventional cooling by heating can
occur for $k\neq0$. Indeed, for $k=1$, Figs. \eqref{Fig7}(c,d),
show that both atom B, with positive temperature, and atom A, with
negative temperature \cite{Note}, cool down by decreasing its internal
energy when their environments are heated up. This effect is saturated
near $n\sim0.5$ for the atom B, which has positive temperature, see
Figs. \eqref{Fig7}(c,d). This effect can be better appreciated by
thinking in the opposite manner, i.e, in the \emph{heating by cooling}:
if the whole environment is cooled down, no matter if the atoms are
with negative or positive temperatures, both atoms always heat up.
This is a remarkable result, since, as emphasized above, it is usually
expected that the energy flux between two systems with opposite signs
to temperatures is from the one with negative to that with positive
temperature, no matter as high the positive temperature is. On the
other hand, the internal energy of atom A, and hence its temperature
(solid black line), always decreases, as expected for systems with
negative temperatures. Actually, energy flux between two systems is
expected to be always from the one with negative to that with positive
temperature\textbf{ }\cite{Frenkel15,Anghel16,Vilar14,Buonsante16}.

It is interesting to note that for $k=2$ and $k=3$, Figs. \eqref{Fig7}(e,f)
and Figs. \eqref{Fig7}(g,h), respectively, the CBH occurs only for
atom A: by increasing the environment temperature, atom B (with positive
temperature) heats up, while atom A (with negative temperature) cools
down. 

\section{Conclusion}

We have proposed an experimentally feasible platform to study systems
capable to display population inversion in its steady state, and thus,
presenting Boltzmann negative temperatures. Our platform includes
i) one qubit and a bosonic mode coupled through an effective Hamiltonian,
the so-called anti-Jaynes-Cummings model (AJCM). The bosonic mode
can be traced out to focus the attention on the qubit, which can display
negative temperatures for a wide range of the cooperativity parameter,
and ii) the previous system in i) plus another qubit, which is coupled
to the first TL system through a natural Hamiltonian model, such as
that stemming from collisions. Using our platform as a theoretical
tool, we were able to study a variety of phenomena, such as the thermalization
for two qubits when one or both of them presents negative temperature,
the control of negative temperatures through the cooperativity parameter,
and also what we have called \emph{unconventional cooling by heating},
which occurs when, decreasing (increasing) the temperatures of the
whole environment, the temperatures of one or both the two qubits
increases (decreases), even when the temperature of one qubit is negative.
This is a striking result if we remember that systems with negative
temperatures are expected to decrease its negative temperature as
in contact with another system having positive temperatures. Our proposal
can be nowadays engineered in several contexts, such as trapped ions
\cite{Leibfried03,Poyatos96}, cavity QED \cite{Rosado15}, nanomechanical
resonator \cite{xue07}, among others \cite{Rempe00,Kimble03,Blatt09}.
\begin{acknowledgments}
We acknowledge financial support from the Brazilian agency CNPq, CAPES
and FAPEG. This work was performed as part of the Brazilian National
Institute of Science and Technology (INCT) for Quantum Information
(Grant No. 465469/2014-0). C.J.V.-B. acknowledges support from Brazilian
agencies No. 2013/04162-5 Sao Paulo Research Foundation (FAPESP) and
from CNPq (Grant No. 308860/2015-2).
\end{acknowledgments}

\end{document}